\documentclass [aps,10pt,prl,twocolumn,longbibliography]{revtex4-2}

\usepackage{graphicx}
\usepackage{subfigure}
\usepackage{amsmath}
\usepackage{amsthm}
\usepackage{amssymb}
\usepackage{hyperref}
\usepackage{xcolor}
\usepackage{textcomp}
\usepackage[normalem]{ulem}
\usepackage{orcidlink}

\hypersetup{colorlinks=true, linkcolor=black, citecolor=black, urlcolor=blue}

\def\be{\begin{equation}}
\def\ee{\end{equation}}

\begin{document}

\title{Nonequilibrium Nonlinear Effects and Dynamical Boson Condensation in a Driven-Dissipative Wannier-Stark Lattice}
 
\author{Arkadiusz Kosior\,\orcidlink{0000-0002-5039-1789}}
\author{Karol Gietka\,\orcidlink{0000-0001-7700-3208}}
\author{Farokh Mivehvar\,\orcidlink{0000-0003-4776-1352}}
\author{Helmut Ritsch\,\orcidlink{0000-0001-7013-5208}}
\affiliation{Institut f\"ur Theoretische Physik, Universit{\"a}t Innsbruck, A-6020~Innsbruck, Austria}


\begin{abstract}
Driven-dissipative light-matter systems can exhibit collective nonequilibrium phenomena due to loss and gain processes on the one hand and effective photon-photon interactions on the other hand. As generic example we study a bosonic lattice system implemented via an array of driven-dissipative coupled nonlinear resonators with linearly increasing resonance frequencies across the lattice. The model also describes a driven-dissipative Bose-Hubbard model in a tilted potential \emph{without} a particle-conservation constraint. We numerically predict a diverse range of stationary and non-stationary states resulting from the interplay of the tilt, tunneling, on-site interactions and loss and gain processes. Our key finding is that, under weak on-site interactions, the bosons mostly condense into a selected, single-particle Wannier-Stark state without exhibiting the expected Bloch oscillations. As the strength of the onsite interactions increase, a non-stationary regime emerges which, surprisingly, exhibits periodic Bloch-type oscillations. As a direct consequence of the driven-dissipative nature of the system we predict a highly nontrivial phase diagram including regular oscillating as well as chaotic dynamical regimes. While a straightforward photonic implementation using microwave or optical modes is possible, such dynamics might also be observable for an ultracold gas in a vertical lattice with gravity or a tilted external potential. 
\end{abstract}

\maketitle



\emph{Introduction.}---The Bose-Hubbard model is a paradigmatic model in condensed matter physics for describing strongly correlated interacting bosons on a lattice \cite{Gersch1963,Dutta2015}. Dissipative behavior in these systems, typically induced by coupling with external degrees of freedom, lead to a range of complex phenomena,  including emergent phase transitions \cite{Miyazaki2003,Vicentini2018,Roberts2023}, pair coherent states \cite{Roberts2023}, dissipation induced correlations \cite{Kiffner2011},  and pair condensation within density-induced tunneling where dissipation can emerge intrinsically \cite{Krzywicka2022,Krzywicka2024}.
Given that the Bose-Hubbard model is well-suited for bosonic systems, could it (or similar models) also be used to describe strongly correlated multi-photon states? 
Although photons do not directly interact with each other, they can effectively interact through their interaction with matter (particularly in non-linear Kerr media), which could give rise to intriguing collective phenomena analogous to condensate matter. Indeed, this research direction was postulated almost 20 years ago \cite{Hartmann2006,Greentree2006} and has been actively developed since then \cite{Noh2016,Hartmann2016}. 

Nevertheless there is a major difference between particle and photon systems: while the matter-particle number is strictly conserved, photons can appear and disappear due to absorption, spontaneous or stimulated emission, and external photon sources. This causes open photonic systems to be inherently out of equilibrium~\cite{Walls2008} and even stationary sates are typically not determined simply by temperature and entropy but rather by the dynamical balance of gain and loss. 
Intriguing nonequilibrium phenomena can thus appear in composite light-matter systems~\cite{Deng2010, Keeling2011, Byrnes2014, Mivehvar2021} and in particular in quantum fluids of light~\cite{Carusotto2013Quantum}. The most notable example is the observation of the quasi-equilibrium Bose-Einstein condensate (BEC) of exciton polaritons---bosonic quasiparticles composed of a mixture of an exciton (an electron-hole pair) and a cavity photon---in a semiconductor microcavity~\cite{Deng2002,Kasprzak2006, Balili2007, Amo2009} and the BEC of photons interacting via molecules in a multimode optical microcavity~\cite{Klaers2010}. Despite the driven-dissipative nature of these systems, they still exhibit an effective thermalization process to which one can attribute an effective temperature. This stands in a sharp contrast to a typical laser operation, where the thermalization is completely ineffective and the photon gas is far out-of-equilibrium.
\begin{figure}[b!]
 \centering
\includegraphics[width=1\columnwidth]{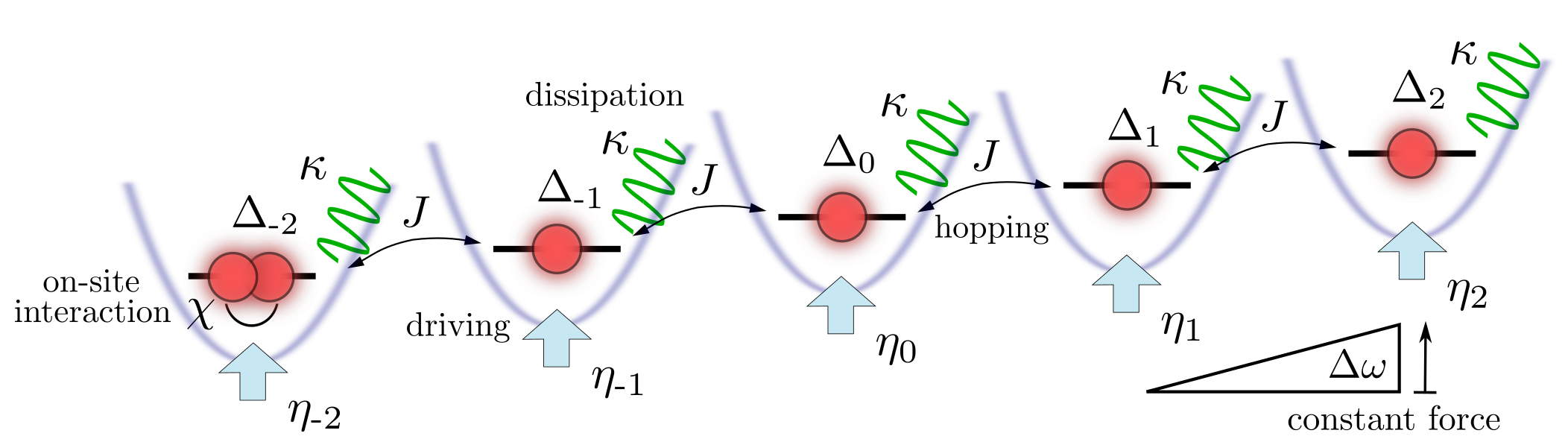}
    \caption{Schematic sketch of the model consisting of an abstract array of driven-dissipative coupled nonlinear resonators  with on-site interaction strength $\chi$ and linearly increasing detunings $\Delta_j$, mimicking a constant force $F=\Delta \omega$. Coherent pumps ($\eta_j$) inject bosons, which can hop between adjacent resonators at a constant rate $J$. Losses $\kappa$ are modeled using the quantum Heisenberg equations \eqref{operator_t}.}\label{model}
\end{figure}

Finally, substantial progress has been achieved in investigating correlated many-body effects with photons~\cite{Noh2016,Hartmann2016}.
In earlier investigations, the focus was on establishing connections between driven-dissipative steady states and equilibrium many-body phases. These included the prediction of a phase transition from a superfluid to a Mott-insulator state for photons via the photon-blockade effect in coupled cavities~\cite{Hartmann2006, Greentree2006,Hartmann2008,Schmidt2009,Tomadin2010,Wu2011,footnote1}.
More recently the focus was shifted towards the intriguing realm of the driven-dissipative regime, where nonequilibrium steady-state phases exhibit distinct properties in comparison to thermally equilibrium cases~\cite{LeBoit2013,LeBoit2014,Grass2019}. For example, the boundary between monostable and bistable phases in a driven-dissipative model resembles characteristic Mott insulator lobes but the mean photon density is not constant within these regions~\cite{LeBoit2013}. In a wider context various ideas and schemes have also been put forward to simulate geometric phases and gauge potentials for photons, opening the possibility for realizing nonequilibrium topological photonic states~\cite{Lu2014, Ozawa2019}.

\begin{figure}[t]
 \centering
\includegraphics[width=1\columnwidth]{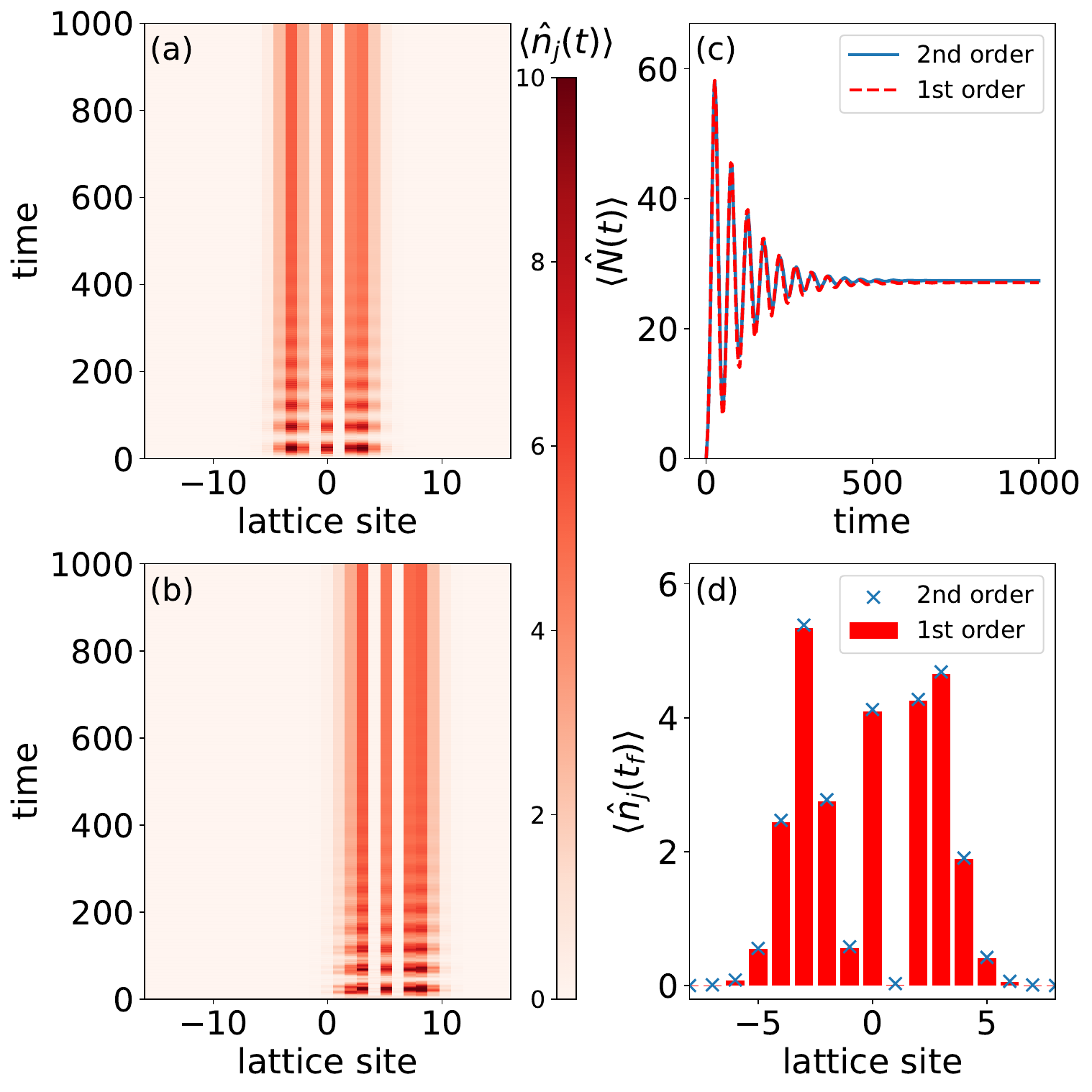}
\caption{Nonequilibrium dynamics of the system for a weak 
on-site
interaction strength $ \chi = 10^{-2}$ and the lattice tilt $\Delta \omega = 0.5$. The system reaches a spatially localized steady state in long time.
The expectation values of the particle number operators $\langle \hat n _j (t) \rangle$ in the course of time evolution in the first-order cumulant expansion (i.e., mean field) for (a) $j_0 = 0$ and (b) $j_0 = 5 $. 
(c)~The time evolution of the expectation value of the total boson
number operator $\langle \hat N \rangle $ for $j_0 = 0$ in both first- and second-order cumulant expansion. (d)~The distribution of $\langle \hat n_j(t_f) \rangle$ over lattice sites in the stationary state for $j_0 = 0$ (cf.\ panel a). To a very good approximation, the distribution is proportional to the probability density of a single WS state (see also Fig.~\ref{fig_bec}), signalling a nonequilibrium Bose-condensation into a WS state. Mean field is quite accurate in this weakly interacting regime. 
}
\label{fig_stationary_states}
\end{figure}

\begin{figure}[t!]
 \centering
\includegraphics[width=1\columnwidth]{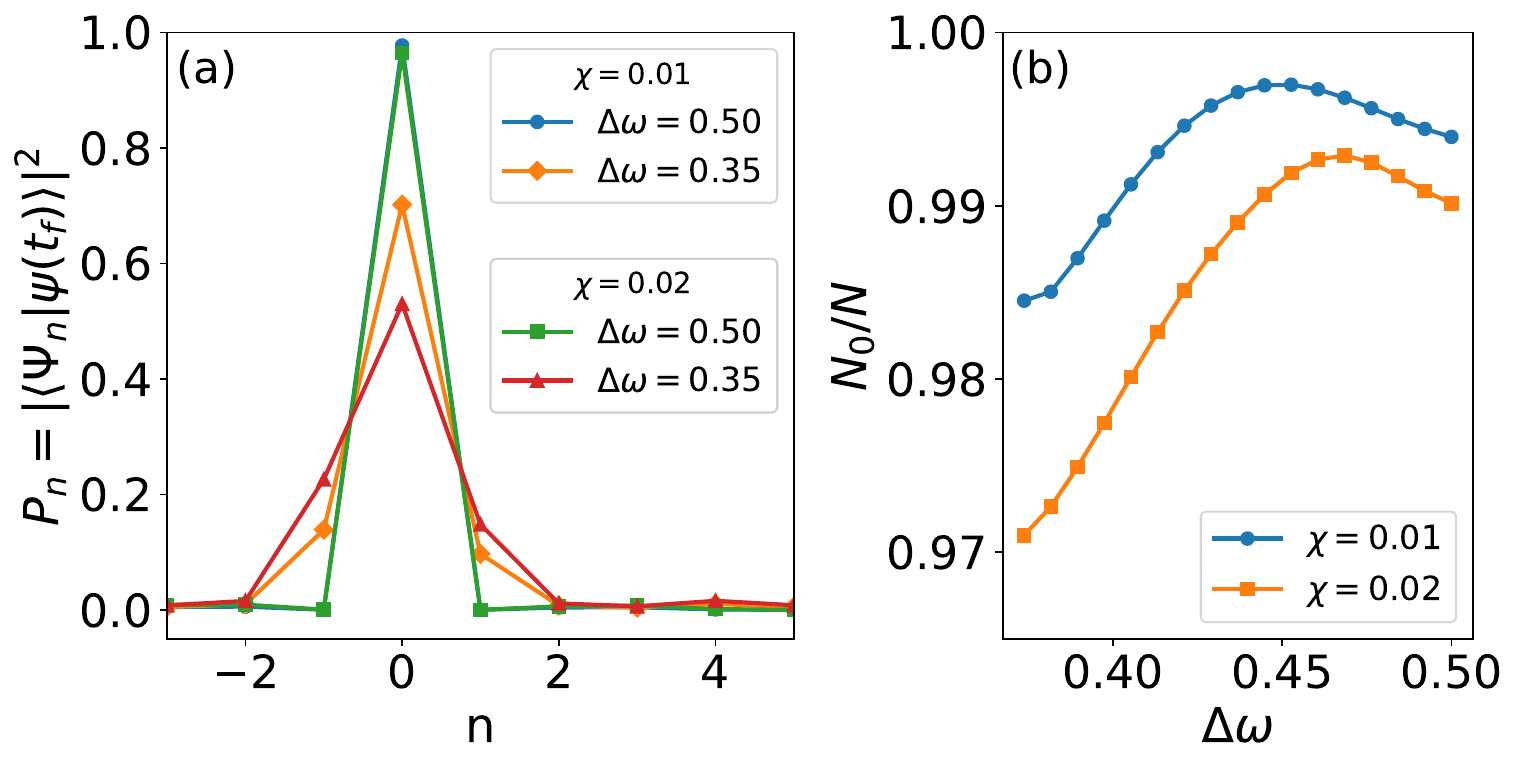}
\caption{
(a) Steady-sate fidelity $P_n(t_f)$ between the mean-field wavefunction $|\psi (t_f) \rangle $ and the WS basis states $|\Psi_n\rangle $. To a very good approximation, the condensate wavefunction is either proportional to only one WS state or is a superposition of a few WS states as in an anti-resonant case (see the discussion in the main text). 
(b)~The
dominant eigenvalue $N_0$ of the single-particle density matrix as a function of $\Delta \omega$ remains close to the total number of bosons $N$, indicating a high condensate fraction.
}\label{fig_bec}
\end{figure}

\begin{figure}[t]
 \centering
\includegraphics[width=1\columnwidth]{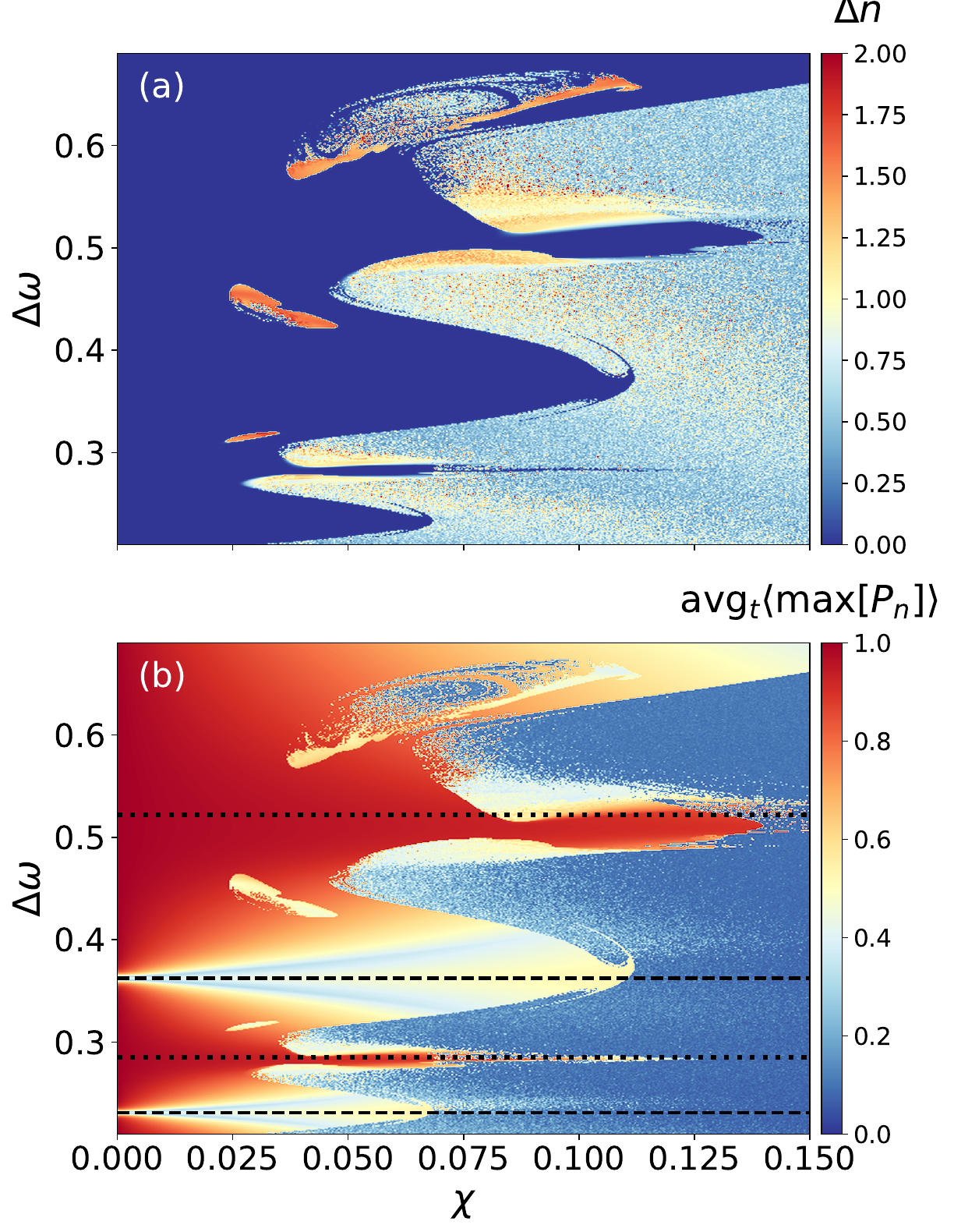}
\caption{Mean-field nonequilibrium many-body phase diagram of the system in the parameter plane of the force $\Delta \omega$ vs.\ the 
on-site
interaction strength $\chi$. 
(a) Relative change of the total 
boson number
over a long-time evolution $\Delta n$ [Eq.~\eqref{delta_n}] reveals three regimes: stationary steady-state phase (deep blue), dynamically unstable chaotic regime (light blue), and non-stationary regular oscillatory states (warm colors). 
(b) Time-averaged maximal fidelity between the mean-field wavefunction $|\psi \rangle $ and the WS basis states $|\Psi_n\rangle $ is complementary to panel (a) and reveals particularly a series of narrow bands that can be explained on a single-particle level as pumping anti-resonances; see the discussion in the main text. The dashed (dotted) lines correspond to zeros of the Bessel function $\mathcal{J}_0$ ($\mathcal{J}_1$). Note that interactions slightly shift the positions of the anti-resonances.
%
%
}\label{fig_phase_diag}
\end{figure}

\begin{figure}[t]
 \centering
\includegraphics[width=1\columnwidth]{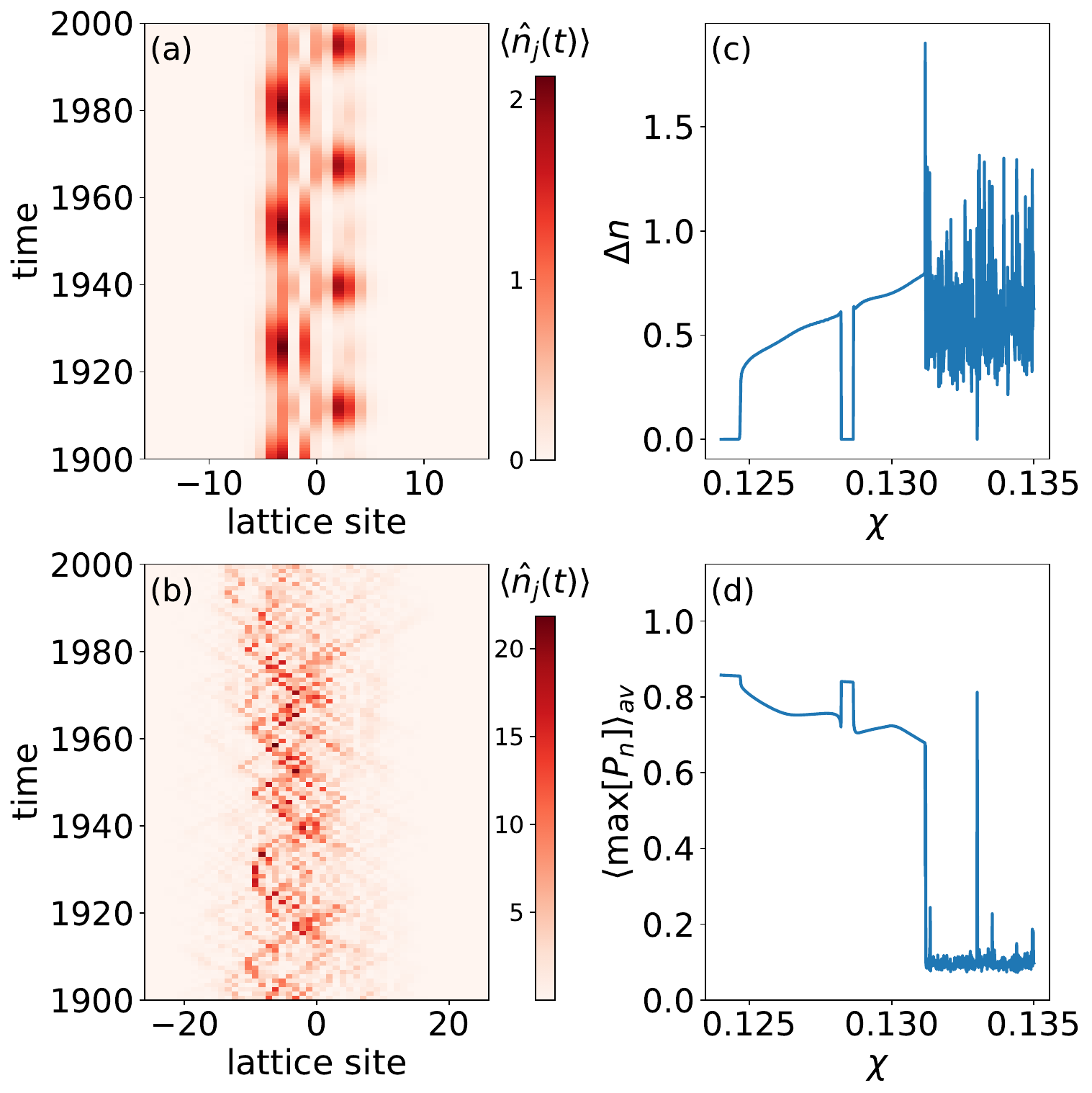}
\caption{Mean-field dynamics of the expectation values of the  number operators $\langle \hat n _j (t) \rangle$ for different interaction strengths: (a) $\chi = 0.13$ in the oscillatory regime, and (b) $\chi = 0.135$ in the chaotic regime. Stationary and different non-stationary solutions are distinguished by monitoring (c) the relative change of the total boson number over a long-time evolution and (d) the time-averaged maximal fidelity. The lattice tilt is set to $\Delta \omega = 0.5$ for all panels.
}\label{fig_bo}
\end{figure}

In this Letter, we investigate a driven-dissipative array of coupled nonlinear resonators with linearly increasing resonant frequencies, see Fig.~\ref{model}. 
In our generic model bosons are continuously injected into  the system. 
Bosons are then redistributed via nearest-neighbor mode couplings until they eventually dissipate (for example, photon leakage through imperfect cavity mirrors). The dynamics of the system can be effectively captured by a driven-dissipative Bose-Hubbard model in a tilted potential \emph{without} particle conservation. Despite the conceptual simplicity, the model, as we demonstrate, exhibits a variety of intriguing stationary and non-stationary, nonequilibrium phenomena controllable by the system parameters. 
A key result of our study is that for sufficiently weak on-site interactions, bosons dynamically condense into a selected, spatially localized state due to explicit $U(1)$ symmetry breaking (Fig.~\ref{fig_stationary_states}), instead of exhibiting the expected Bloch oscillations \cite{Wannier1960,Hartmann2004}. Specifically, the condensate wavefunction is often close to a single Wannier-Stark (WS) state, with only small contributions from neighboring WS states (Fig.~\ref{fig_bec}). Interestingly, increasing the strength of local on-site interactions drives the system into a non-stationary regime (Fig.~\ref{fig_phase_diag}), where the bosonic density undergoes periodic Bloch-type oscillations over time which are induced by interactions (Fig.~\ref{fig_bo}) \cite{Peschel1998,Sapienza2003,Yuan2016}. This contrasts sharply with the irreversible decay of Bloch oscillations in interacting atoms within a 1D tilted lattice \cite{Buchleitner2003,Gustavsson2008,Krimer2009,Eckstein2011,footnote2}. Notably, both regimes are independent of initial conditions and the choice of the pumped resonator, stemming directly from the driven-dissipative nature of the system. The generic nature of our model suggests that experimental realization is feasible on various platforms, including superconducting circuits \cite{Houck2012,Jin2013,Jin2014}, photonic crystal structures \cite{Majumdar2012}, waveguide-coupled optical cavities \cite{Lepert2011}, coupled photonic microcavities \cite{Xiao2008,Rodriguez2016},  exciton polariton lattices \cite{Schneider2016}, cold atoms coupled to photonic crystals \cite{Douglas2015}
 and atom-filled transverse multi-mode cavities \cite{Kollr2015}.
Although photonic implementation seems most promising, we note that such dynamics might also be observable for an
ultracold gas in a vertical lattice with gravity or a titled optical potential \cite{Zelan2010,Sharma2021}.


\emph{Model and its Hamiltonian.}---Consider an array of coupled resonators (labeled by $j \in \mathbb{Z}$) with linearly increasing resonant frequencies $\omega_j \propto j$, each containing a Kerr-like non-linear medium. Coherent pumps with the frequency $\omega_p$ continuously inject bosons into the resonator modes. Each resonator is coupled to two adjacent resonators
which leads to a coherent hopping of bosons in the 
resonator lattice.
The Hamiltonian of the system is given by~\cite{SM}, 
\begin{align}\label{hamiltonian}
\hat H &= \sum_j \left[ \hbar \Delta_j \hat a_j^\dagger \hat a_j - J \big(\hat a_j^\dagger \hat a_{j+1} + \text{H.c.} \big) + \chi \hat a_j^{\dagger2} \hat a_j^2 \right]
\nonumber \\
&+ \hbar \sum_j \eta_j \left(\hat a_j + \hat a_j^\dagger \right), 
\end{align}
with $\hat a_j$, $\hat a_j^\dagger$ being bosonic operators annihilating and creating a boson in the $j$-th resonator, respectively. Here we have defined
\be\label{Delta_n}
\Delta_j = \omega_j - \omega_p \equiv \Delta \omega \left( j - j_0 \right),
\ee
as the resonator-pump detuning. Moreover, $J$ is the nearest-neighbor tunneling-amplitude rate, $\chi$ the on-site 
interaction strength due to the effective Kerr non-linearity, and $\eta_j$ the pumping rate of the $j$-th resonator. Bosonic losses, assumed to be the same throughout the lattice, $\kappa$ are taken into account via 
the quantum Heisenberg equations of motion,
\be\label{operator_t}
\frac{d \hat{a}_j }{dt} = \frac{i}{\hbar}[ \hat H, \hat{a}_j ] -\kappa \hat{a}_j. 
\ee

The Hamiltonian \eqref{hamiltonian} is an effective time-independent Hamiltonian expressed in the rotating frame of the coherent pumps, which is quite general and can be applied to many experimental scenarios (for example, such a  Hamiltonian can be designed within cicuit-QED setup \cite{Jin2013}, but is also an effective lowest Bloch band  Hamiltonian of ultracold atoms in driven dissipative optical lattices \cite{Sharma2021}, see also \cite{Noh2016,Hartmann2016}).
The first line of the Hamiltonian \eqref{hamiltonian} describes the familiar equilibrium Bose-Hubbard model in a tilted lattice. While in the non-interacting limit the equilibrium model features well-known Bloch oscillations \cite{Wannier1960,Hartmann2004}, it has been recently shown that strong interactions can lead to disorder-free many-body localization~\cite{Yao2020,Taylor2020} (for related experiments see Ref.~\cite{Morong2021}). The second line of the Hamiltonian~\eqref{hamiltonian} introduces a coherent pumping, which along with the  environment decay $\kappa$ [see Eq.~\eqref{operator_t}] explicitly breaks the $U(1)$ symmetry of the system associated with the particle number conservation~\cite{SM}. In the following we will show that this lack of the particle conservation and explicitly broken $U(1)$ symmetry due to the loss and gain processes has fundamental consequences in both static and dynamics of the system.





\emph{Consequences of explicit U(1) symmetry breaking: State selection}.---In order to gain some physical intuition,  let us start with the the non-interacting limit, $\chi=0$. In the WS basis, the Heisenberg equations of motions read~\cite{SM}, 
\be \label{eq:WS-eq-motion-b}
 i \frac{d\hat b_n}{dt} = \left( \Delta_n- i \kappa \right) \hat b_n + \tilde \eta_n, 
\ee
where $ \hat b_n = \sum_j \beta_{n,j} \hat a_n$, $\tilde \eta_n = \sum_j \beta_{n,j} \eta_j$, and $\beta_{n,j} = \mathcal{J}_{j-n}\left({2 J}/{\Delta \omega}\right) $, with $\mathcal{J}_k$ being the Bessel function of the first kind of order $k$. 
The equations of motion~\eqref{eq:WS-eq-motion-b} readily yield
\be\label{bn}
\hat b_n(t) = e^{-i t( \Delta_n -i \kappa)} \,\hat b_n(0) + \frac{e^{-i t( \Delta_n -i \kappa)}-1}{\Delta_n -i \kappa} \, \tilde \eta_n.
\ee
If $\forall_n \tilde{\eta}_n = 0$, the solutions correspond to damped Bloch oscillations \cite{SM}. If additionally $\kappa = 0$, one then recovers the common Bloch oscillations and the choice of $j_0$ in Eq.~\eqref{Delta_n} becomes arbitrary as it only adds an irrelevant phase factor. 

In the long-time limit $ t \gg \kappa^{-1} \gg J^{-1}$, regardless of the choice of initial conditions $\hat b_n(0)$, the Bloch oscillations are completely damped out and the system reaches a steady state with the WS mode occupations,  
\be \label{eq:ss-WS-mode-occupation}
\langle \hat n_n \rangle = \frac{\tilde \eta_n^2}{(\Delta \omega)^2 (n-j_0)^2 + \kappa^2}.
\ee
As can be seen from Eq.~\eqref{eq:ss-WS-mode-occupation}, now $j_0$ plays an essential role. In particular, by properly choosing $j_0\in \mathbb{Z}$  one can select dynamically a single WS mode ($ n = j_0$) to be microscopically occupied. However, we note this approach does not work when $\tilde \eta_{j_0} = 0$, which happens for some specific ratios of $J/\Delta \omega$. In these cases, one encounters a sequence of pumping anti-resonances, which lead to the occupation a few adjacent WS states instead. Furthermore, these anti-resonances are responsible for non-trivial phase boundaries between stationary and non-stationary states as shown in Fig.~\ref{fig_phase_diag}, which we will delve into it in the last section.




The above simple analysis is valid qualitatively also for sufficiently small 
on-site
interactions, which we confirm numerically in the following. In numerical simulations we set $ J = \hbar = 1$ (as the unit of energy), $\kappa = 10^{-2}$, $\forall_j \hat a_j(0) = 0$ 
, and calculate the expectation values of the relevant operators using the cumulant expansion in both first (i.e., mean-field) and second order~\cite{Plankensteiner2022,SM}. Moreover, for the sake of simplicity and without loss of generality, we consider only a single resonator pumping $\eta_j = \eta\delta_{j,0}$, but different choices do not change our main conclusions~\cite{SM}. Finally, with the exception of Fig.~\ref{fig_stationary_states}, we also fix $j_0 = 0$. 



\emph{Nonequilibrium condensation in the weakly interacting regime.}---Now, we turn our attention to the weakly interacting regime, $\chi\ll1$. As in the non-interacting case, in the weakly interacting regime the many-body bosonic system still  occupies macroscopically a single or a few one-particle WS states with a high degree of coherence. In Fig.~\ref{fig_stationary_states}, we show the expectation value of number operators $\hat n _j = \hat a_j^\dagger \hat a_j $ as well as the 
total boson-number operator $\hat N = \sum_j \hat n_j$. As can be seen, in each case the system reaches a spatially-localised stationary steady state which, for the chosen parameters, is proportional to a single WS state with $n=j_0$.

To further quantify this observation, we calculate the fidelity 
$P_n(t) = | \langle \Psi_n | \psi (t) \rangle |^2$ 
between the mean-field wavefunction $|\psi (t) \rangle $ and the WS basis states $|\Psi_n\rangle $. The distribution of $P_n$ is illustrated in Fig.~\ref{fig_bec}(a) in steady states, showing it is centered around $n=j_0 = 0$ [cf.\ Eq.~\eqref{eq:ss-WS-mode-occupation}]. 
In general two distinct scenarios are possible: (i) The mean-field wavefunction consists predominantly of a single WS state, or (ii) it has contributions from a few different WS states. As Fig.~\ref{fig_bec}(a) shows, for $\Delta \omega = 0.5$ [as in Fig.~\ref{fig_stationary_states}(a) and (d)] the wavefunction is close to the central WS state, while for $\Delta \omega = 0.35$ two additional modes $n=\pm 1$ are also significantly occupied at the macroscopic level.
This is because of the aforementioned pumping anti-resonances, where the population of $n = \pm 1$ ($n = 0$) mode is completely suppressed at $\Delta \omega \approx 0.522$ ($\Delta \omega \approx 0.362$) due to the hitting a zero of the $\mathcal{J}_{1}$ ($\mathcal{J}_{0}$) Bessel function. 

In order to confirm that we deal with Bose condensation, 
we consider the single-particle reduced density matrix $\hat \rho_1(x,x') = \langle \hat{\psi}^\dagger(x) \hat{\psi}(x') \rangle$, whose eigenvalues determine the occupation probabilities of the natural orbitals~\cite{Pethick2008}. The highest occupation number quantifies the level of coherence in the system. Although we can expand the field operators in any orthogonal basis, we choose the Wannier basis $\hat \psi (x) = \sum_j w_j(x) \hat a_j$ and calculate the eigenvalues of $\langle \hat a^\dagger_j \hat a_l \rangle $ in a steady state of the system in the second-order cumulant expansion~\cite{SM}.
Indeed, as can be seen from Fig.~\ref{fig_bec}(b) the highest eigenvalue $N_0$ is very close to the total number of  bosons $N = \langle \hat N \rangle $, supporting the interpretation of a dynamical BEC. In contrast to the distribution of $P_n$ as shown in Fig.~\ref{fig_bec}(a), the highest eigenvalue of the the reduced density matrix is only weakly affected by the lattice tilt $\Delta \omega$. Hence, we infer that the condensate wavefunction is either close to a single WS states or is a superposition of a few WS states. 


\emph{Non-stationary phases in the strongly interacting regime.}---Above we showed that in the non-interacting and weakly interacting regimes bosons can condensate into a one or a few selected WS states and reach a steady state. However, with increasing the interaction strength $\chi$, the stationary states lose their stability and intriguing non-stationary solutions appear. Figure~\ref{fig_phase_diag}(a) depicts the nonequilibrium many-body phase diagram of the system in the parameter plane of $\{\chi,\Delta\omega\}$ and contains three main regimes: stationary steady states (deep blue), dynamically unstable chaotic regime (light blue), and non-stationary regular oscillatory states (warm colors). The three phases are distinguished by the relative change of the total boson number over a long-time evolution, defined as 
\be\label{delta_n}
\Delta n = \frac{\max_t \langle \hat N(t) \rangle - \min_t \langle \hat N(t) \rangle}{ \text{avg}_t \langle \hat N (t) \rangle},
\ee
with $\max_t $, $\min_t$, and $\text{avg}_t$ denoting, respectively, the maximal, minimal, and average value of a quantum-averaged observable during a long-time evolution (where the initial transient dynamics of the system has been neglected). 
Although the phase diagram is dominated by the stationary and dynamically unstable chaotic regimes, regions of regular oscillatory dynamics appear mostly on boundaries between the two regimes; see also Fig.~\ref{fig_phase_diag}(b) showing the time-averaged maximal fidelity and Supplemental Material~\cite{SM}. 

Interestingly, bosons in the regular oscillatory regime tunnel between a few neighboring lattice sites performing spatially-confined oscillations reminiscent of standard Bloch oscillations; see Fig.~\ref{fig_bo}. Unlike standard Bloch oscillations that tend to decay in interacting systems~\cite{Buchleitner2003,Gustavsson2008,Krimer2009,Eckstein2011,footnote2}, in our driven-dissipative model oscillatory solutions are in fact induced by the on-site interactions and persist over long time dynamics, even beyond the mean-field regime, although with quantitative differences~\cite{SM}, as expected from literature \cite{Polak2005,Huybrechts2020}. These solutions can be conceptualized as multimode limit cycles (as shown in Supplemental Material~\cite{SM}) within driven-dissipative systems, where the steady-state solutions of equations of motion exhibit dynamic instability \cite{Piazza2015, Chan2015, Owen2018, Kessler2019, Colella2022, Gao2023, Kosior2023b, Mivehvar2024, Skulte2024}.




\emph{Summary, conclusions, and perspectives.}---
In summary, we studied an array of coupled nonlinear resonators with linearly increasing resonant frequencies, which is equivalent to a $U(1)$-symmetry broken driven-dissipative Bose-Hubbard model with a tilted potential. 
It could be equally implemented based on frequency equidistant weakly coupled modes of a single resonator. Our semi-analytical analysis reveals a range of both stationary and non-stationary nonequilibrium phenomena. Notably, bosons condense into selected, stationary WS states under sufficiently weak on-site interactions. As the strength of 
interactions increases, we observe a transition to a non-stationary dynamical regime marked by periodic Bloch-like oscillations over time. Unlike the equilibrium counterpart, these regular oscillations are induced by the 
interactions and do not decay over time. 

Our research sheds light on the intricate behavior of driven-dissipative coupled resonator systems, emphasizing the role of explicit $U(1)$ symmetry breaking and interactions in shaping their dynamics. These findings hold promise for applications such as coherent light storage~\cite{storage2001atomicvapor,lightstorage2013minute,lightstorage2018vapor}, light confinement~\cite{confinement2000randommedia,confinement2011periodicity,confinement2014disorededmedia}, generation of non-classical many-photon states~\cite{Cooper:13fockstates,manybody2014luking,emission2023manybody}, and distributed quantum sensing~\cite{distriubtesensing2020network,Zhang_2021,gietka2023opticallatt}.
While our focus in this Letter has been on weakly interacting systems, we underscore the importance of delving deeper into dynamics beyond perturbation regimes. Specifically, it is interesting to explore the non-stationary states in many-body regimes, particularly in the context of ergodicity breaking~\cite{DeLuca_2013ergodi,Gietka2019ergodic} and Stark many-body localization~\cite{STARKmbl2019pollmann}. Another intriguing scenario is to extend our driven-dissipative model to topological setting, where exotic nonequilibrium topological effects are expected to appear~\cite{Hafezi2013, Mittal2014, Gong2018, Colella2019}. 

The data presented in this article is available from \cite{zenodo}.

\begin{acknowledgments}
We acknowledge stimulating discussions with Darrick Chang and Thomas Pohl. 
This research was funded in whole or in part by the Austrian Science Fund (FWF) [grant DOIs: 10.55776/ESP171, 10.55776/M3304 and 10.55776/P35891]. For open access purposes, the authors
have applied a CC BY public copyright license to any author accepted manuscript version arising from this submission. F.\,M. additionally acknowledges the support of the ESQ-Discovery Grant of the Austrian Academy of Sciences (\"OAW). 
\end{acknowledgments}


\begin{thebibliography}{100}%
\makeatletter
\providecommand \@ifxundefined [1]{%
 \@ifx{#1\undefined}
}%
\providecommand \@ifnum [1]{%
 \ifnum #1\expandafter \@firstoftwo
 \else \expandafter \@secondoftwo
 \fi
}%
\providecommand \@ifx [1]{%
 \ifx #1\expandafter \@firstoftwo
 \else \expandafter \@secondoftwo
 \fi
}%
\providecommand \natexlab [1]{#1}%
\providecommand \enquote  [1]{``#1''}%
\providecommand \bibnamefont  [1]{#1}%
\providecommand \bibfnamefont [1]{#1}%
\providecommand \citenamefont [1]{#1}%
\providecommand \href@noop [0]{\@secondoftwo}%
\providecommand \href [0]{\begingroup \@sanitize@url \@href}%
\providecommand \@href[1]{\@@startlink{#1}\@@href}%
\providecommand \@@href[1]{\endgroup#1\@@endlink}%
\providecommand \@sanitize@url [0]{\catcode `\\12\catcode `\$12\catcode
  `\&12\catcode `\#12\catcode `\^12\catcode `\_12\catcode `\%12\relax}%
\providecommand \@@startlink[1]{}%
\providecommand \@@endlink[0]{}%
\providecommand \url  [0]{\begingroup\@sanitize@url \@url }%
\providecommand \@url [1]{\endgroup\@href {#1}{\urlprefix }}%
\providecommand \urlprefix  [0]{URL }%
\providecommand \Eprint [0]{\href }%
\providecommand \doibase [0]{https://doi.org/}%
\providecommand \selectlanguage [0]{\@gobble}%
\providecommand \bibinfo  [0]{\@secondoftwo}%
\providecommand \bibfield  [0]{\@secondoftwo}%
\providecommand \translation [1]{[#1]}%
\providecommand \BibitemOpen [0]{}%
\providecommand \bibitemStop [0]{}%
\providecommand \bibitemNoStop [0]{.\EOS\space}%
\providecommand \EOS [0]{\spacefactor3000\relax}%
\providecommand \BibitemShut  [1]{\csname bibitem#1\endcsname}%
\let\auto@bib@innerbib\@empty
\bibitem [{\citenamefont {Gersch}\ and\ \citenamefont
  {Knollman}(1963)}]{Gersch1963}%
  \BibitemOpen
  \bibfield  {author} {\bibinfo {author} {\bibfnamefont {H.~A.}\ \bibnamefont
  {Gersch}}\ and\ \bibinfo {author} {\bibfnamefont {G.~C.}\ \bibnamefont
  {Knollman}},\ }\bibfield  {title} {\bibinfo {title} {Quantum cell model for
  bosons},\ }\href {https://doi.org/10.1103/PhysRev.129.959} {\bibfield
  {journal} {\bibinfo  {journal} {Phys. Rev.}\ }\textbf {\bibinfo {volume}
  {129}},\ \bibinfo {pages} {959} (\bibinfo {year} {1963})}\BibitemShut
  {NoStop}%
\bibitem [{\citenamefont {Dutta}\ \emph {et~al.}(2015)\citenamefont {Dutta},
  \citenamefont {Gajda}, \citenamefont {Hauke}, \citenamefont {Lewenstein},
  \citenamefont {L\"{u}hmann}, \citenamefont {Malomed}, \citenamefont
  {Sowiński},\ and\ \citenamefont {Zakrzewski}}]{Dutta2015}%
  \BibitemOpen
  \bibfield  {author} {\bibinfo {author} {\bibfnamefont {O.}~\bibnamefont
  {Dutta}}, \bibinfo {author} {\bibfnamefont {M.}~\bibnamefont {Gajda}},
  \bibinfo {author} {\bibfnamefont {P.}~\bibnamefont {Hauke}}, \bibinfo
  {author} {\bibfnamefont {M.}~\bibnamefont {Lewenstein}}, \bibinfo {author}
  {\bibfnamefont {D.-S.}\ \bibnamefont {L\"{u}hmann}}, \bibinfo {author}
  {\bibfnamefont {B.~A.}\ \bibnamefont {Malomed}}, \bibinfo {author}
  {\bibfnamefont {T.}~\bibnamefont {Sowiński}},\ and\ \bibinfo {author}
  {\bibfnamefont {J.}~\bibnamefont {Zakrzewski}},\ }\bibfield  {title}
  {\bibinfo {title} {Non-standard hubbard models in optical lattices: a
  review},\ }\href {https://doi.org/10.1088/0034-4885/78/6/066001} {\bibfield
  {journal} {\bibinfo  {journal} {Reports on Progress in Physics}\ }\textbf
  {\bibinfo {volume} {78}},\ \bibinfo {pages} {066001} (\bibinfo {year}
  {2015})}\BibitemShut {NoStop}%
\bibitem [{\citenamefont {Miyazaki}\ \emph {et~al.}(2003)\citenamefont
  {Miyazaki}, \citenamefont {Takahide}, \citenamefont {Kanda},\ and\
  \citenamefont {Ootuka}}]{Miyazaki2003}%
  \BibitemOpen
  \bibfield  {author} {\bibinfo {author} {\bibfnamefont {H.}~\bibnamefont
  {Miyazaki}}, \bibinfo {author} {\bibfnamefont {Y.}~\bibnamefont {Takahide}},
  \bibinfo {author} {\bibfnamefont {A.}~\bibnamefont {Kanda}},\ and\ \bibinfo
  {author} {\bibfnamefont {Y.}~\bibnamefont {Ootuka}},\ }\bibfield  {title}
  {\bibinfo {title} {Quantum fluctuations and dissipative phase transition in
  one-dimensional josephson junction arrays},\ }\href
  {https://doi.org/10.1016/s1386-9477(02)00949-9} {\bibfield  {journal}
  {\bibinfo  {journal} {Physica E: Low-dimensional Systems and Nanostructures}\
  }\textbf {\bibinfo {volume} {18}},\ \bibinfo {pages} {41–42} (\bibinfo
  {year} {2003})}\BibitemShut {NoStop}%
\bibitem [{\citenamefont {Vicentini}\ \emph {et~al.}(2018)\citenamefont
  {Vicentini}, \citenamefont {Minganti}, \citenamefont {Rota}, \citenamefont
  {Orso},\ and\ \citenamefont {Ciuti}}]{Vicentini2018}%
  \BibitemOpen
  \bibfield  {author} {\bibinfo {author} {\bibfnamefont {F.}~\bibnamefont
  {Vicentini}}, \bibinfo {author} {\bibfnamefont {F.}~\bibnamefont {Minganti}},
  \bibinfo {author} {\bibfnamefont {R.}~\bibnamefont {Rota}}, \bibinfo {author}
  {\bibfnamefont {G.}~\bibnamefont {Orso}},\ and\ \bibinfo {author}
  {\bibfnamefont {C.}~\bibnamefont {Ciuti}},\ }\bibfield  {title} {\bibinfo
  {title} {Critical slowing down in driven-dissipative bose-hubbard lattices},\
  }\href {https://doi.org/10.1103/PhysRevA.97.013853} {\bibfield  {journal}
  {\bibinfo  {journal} {Phys. Rev. A}\ }\textbf {\bibinfo {volume} {97}},\
  \bibinfo {pages} {013853} (\bibinfo {year} {2018})}\BibitemShut {NoStop}%
\bibitem [{\citenamefont {Roberts}\ and\ \citenamefont
  {Clerk}(2023)}]{Roberts2023}%
  \BibitemOpen
  \bibfield  {author} {\bibinfo {author} {\bibfnamefont {D.}~\bibnamefont
  {Roberts}}\ and\ \bibinfo {author} {\bibfnamefont {A.~A.}\ \bibnamefont
  {Clerk}},\ }\bibfield  {title} {\bibinfo {title} {Competition between
  two-photon driving, dissipation, and interactions in bosonic lattice models:
  An exact solution},\ }\href {https://doi.org/10.1103/PhysRevLett.130.063601}
  {\bibfield  {journal} {\bibinfo  {journal} {Phys. Rev. Lett.}\ }\textbf
  {\bibinfo {volume} {130}},\ \bibinfo {pages} {063601} (\bibinfo {year}
  {2023})}\BibitemShut {NoStop}%
\bibitem [{\citenamefont {Kiffner}\ and\ \citenamefont
  {Hartmann}(2011)}]{Kiffner2011}%
  \BibitemOpen
  \bibfield  {author} {\bibinfo {author} {\bibfnamefont {M.}~\bibnamefont
  {Kiffner}}\ and\ \bibinfo {author} {\bibfnamefont {M.~J.}\ \bibnamefont
  {Hartmann}},\ }\bibfield  {title} {\bibinfo {title} {Dissipation-induced
  correlations in one-dimensional bosonic systems},\ }\href
  {https://doi.org/10.1088/1367-2630/13/5/053027} {\bibfield  {journal}
  {\bibinfo  {journal} {New Journal of Physics}\ }\textbf {\bibinfo {volume}
  {13}},\ \bibinfo {pages} {053027} (\bibinfo {year} {2011})}\BibitemShut
  {NoStop}%
\bibitem [{\citenamefont {Krzywicka}\ and\ \citenamefont
  {Polak}(2022)}]{Krzywicka2022}%
  \BibitemOpen
  \bibfield  {author} {\bibinfo {author} {\bibfnamefont {A.}~\bibnamefont
  {Krzywicka}}\ and\ \bibinfo {author} {\bibfnamefont {T.}~\bibnamefont
  {Polak}},\ }\bibfield  {title} {\bibinfo {title} {Coexistence of two kinds of
  superfluidity at finite temperatures in optical lattices},\ }\href
  {https://doi.org/10.1016/j.aop.2022.168973} {\bibfield  {journal} {\bibinfo
  {journal} {Annals of Physics}\ }\textbf {\bibinfo {volume} {443}},\ \bibinfo
  {pages} {168973} (\bibinfo {year} {2022})}\BibitemShut {NoStop}%
\bibitem [{\citenamefont {Krzywicka}\ and\ \citenamefont
  {Polak}(2024)}]{Krzywicka2024}%
  \BibitemOpen
  \bibfield  {author} {\bibinfo {author} {\bibfnamefont {A.}~\bibnamefont
  {Krzywicka}}\ and\ \bibinfo {author} {\bibfnamefont {T.~P.}\ \bibnamefont
  {Polak}},\ }\bibfield  {title} {\bibinfo {title} {Reentrant phase behavior in
  systems with density-induced tunneling},\ }\bibfield  {journal} {\bibinfo
  {journal} {Scientific Reports}\ }\textbf {\bibinfo {volume} {14}},\ \href
  {https://doi.org/10.1038/s41598-024-60955-1} {10.1038/s41598-024-60955-1}
  (\bibinfo {year} {2024})\BibitemShut {NoStop}%
\bibitem [{\citenamefont {Hartmann}\ \emph {et~al.}(2006)\citenamefont
  {Hartmann}, \citenamefont {Brandão},\ and\ \citenamefont
  {Plenio}}]{Hartmann2006}%
  \BibitemOpen
  \bibfield  {author} {\bibinfo {author} {\bibfnamefont {M.~J.}\ \bibnamefont
  {Hartmann}}, \bibinfo {author} {\bibfnamefont {F.~G. S.~L.}\ \bibnamefont
  {Brandão}},\ and\ \bibinfo {author} {\bibfnamefont {M.~B.}\ \bibnamefont
  {Plenio}},\ }\bibfield  {title} {\bibinfo {title} {Strongly interacting
  polaritons in coupled arrays of cavities},\ }\href
  {https://doi.org/10.1038/nphys462} {\bibfield  {journal} {\bibinfo  {journal}
  {Nat. Phys}\ }\textbf {\bibinfo {volume} {2}},\ \bibinfo {pages} {849–855}
  (\bibinfo {year} {2006})}\BibitemShut {NoStop}%
\bibitem [{\citenamefont {Greentree}\ \emph {et~al.}(2006)\citenamefont
  {Greentree}, \citenamefont {Tahan}, \citenamefont {Cole},\ and\ \citenamefont
  {Hollenberg}}]{Greentree2006}%
  \BibitemOpen
  \bibfield  {author} {\bibinfo {author} {\bibfnamefont {A.~D.}\ \bibnamefont
  {Greentree}}, \bibinfo {author} {\bibfnamefont {C.}~\bibnamefont {Tahan}},
  \bibinfo {author} {\bibfnamefont {J.~H.}\ \bibnamefont {Cole}},\ and\
  \bibinfo {author} {\bibfnamefont {L.~C.~L.}\ \bibnamefont {Hollenberg}},\
  }\bibfield  {title} {\bibinfo {title} {Quantum phase transitions of light},\
  }\href {https://doi.org/10.1038/nphys466} {\bibfield  {journal} {\bibinfo
  {journal} {Nat. Phys}\ }\textbf {\bibinfo {volume} {2}},\ \bibinfo {pages}
  {856–861} (\bibinfo {year} {2006})}\BibitemShut {NoStop}%
\bibitem [{\citenamefont {Noh}\ and\ \citenamefont
  {Angelakis}(2016)}]{Noh2016}%
  \BibitemOpen
  \bibfield  {author} {\bibinfo {author} {\bibfnamefont {C.}~\bibnamefont
  {Noh}}\ and\ \bibinfo {author} {\bibfnamefont {D.~G.}\ \bibnamefont
  {Angelakis}},\ }\bibfield  {title} {\bibinfo {title} {Quantum simulations and
  many-body physics with light},\ }\href
  {https://doi.org/10.1088/0034-4885/80/1/016401} {\bibfield  {journal}
  {\bibinfo  {journal} {Rep. Prog. Phys.}\ }\textbf {\bibinfo {volume} {80}},\
  \bibinfo {pages} {016401} (\bibinfo {year} {2016})}\BibitemShut {NoStop}%
\bibitem [{\citenamefont {Hartmann}(2016)}]{Hartmann2016}%
  \BibitemOpen
  \bibfield  {author} {\bibinfo {author} {\bibfnamefont {M.~J.}\ \bibnamefont
  {Hartmann}},\ }\bibfield  {title} {\bibinfo {title} {Quantum simulation with
  interacting photons},\ }\href
  {https://doi.org/10.1088/2040-8978/18/10/104005} {\bibfield  {journal}
  {\bibinfo  {journal} {J. Opt.}\ }\textbf {\bibinfo {volume} {18}},\ \bibinfo
  {pages} {104005} (\bibinfo {year} {2016})}\BibitemShut {NoStop}%
\bibitem [{\citenamefont {Walls}\ and\ \citenamefont
  {Milburn}(2008)}]{Walls2008}%
  \BibitemOpen
  \bibfield  {author} {\bibinfo {author} {\bibfnamefont {D.~F.}\ \bibnamefont
  {Walls}}\ and\ \bibinfo {author} {\bibfnamefont {G.~F.}\ \bibnamefont
  {Milburn}},\ }\href {https://doi.org/10.1007/978-3-540-28574-8} {\emph
  {\bibinfo {title} {Quantum {O}ptics}}}\ (\bibinfo  {publisher} {Springer
  Berlin Heidelberg},\ \bibinfo {year} {2008})\BibitemShut {NoStop}%
\bibitem [{\citenamefont {Deng}\ \emph {et~al.}(2010)\citenamefont {Deng},
  \citenamefont {Haug},\ and\ \citenamefont {Yamamoto}}]{Deng2010}%
  \BibitemOpen
  \bibfield  {author} {\bibinfo {author} {\bibfnamefont {H.}~\bibnamefont
  {Deng}}, \bibinfo {author} {\bibfnamefont {H.}~\bibnamefont {Haug}},\ and\
  \bibinfo {author} {\bibfnamefont {Y.}~\bibnamefont {Yamamoto}},\ }\bibfield
  {title} {\bibinfo {title} {Exciton-polariton {B}ose-{E}instein
  condensation},\ }\href {https://doi.org/10.1103/RevModPhys.82.1489}
  {\bibfield  {journal} {\bibinfo  {journal} {Rev. Mod. Phys.}\ }\textbf
  {\bibinfo {volume} {82}},\ \bibinfo {pages} {1489} (\bibinfo {year}
  {2010})}\BibitemShut {NoStop}%
\bibitem [{\citenamefont {Keeling}\ and\ \citenamefont
  {Berloff}(2011)}]{Keeling2011}%
  \BibitemOpen
  \bibfield  {author} {\bibinfo {author} {\bibfnamefont {J.}~\bibnamefont
  {Keeling}}\ and\ \bibinfo {author} {\bibfnamefont {N.~G.}\ \bibnamefont
  {Berloff}},\ }\bibfield  {title} {\bibinfo {title} {Exciton–polariton
  condensation},\ }\href {https://doi.org/10.1080/00107514.2010.550120}
  {\bibfield  {journal} {\bibinfo  {journal} {Contemp. Phys.}\ }\textbf
  {\bibinfo {volume} {52}},\ \bibinfo {pages} {131–151} (\bibinfo {year}
  {2011})}\BibitemShut {NoStop}%
\bibitem [{\citenamefont {Byrnes}\ \emph {et~al.}(2014)\citenamefont {Byrnes},
  \citenamefont {Kim},\ and\ \citenamefont {Yamamoto}}]{Byrnes2014}%
  \BibitemOpen
  \bibfield  {author} {\bibinfo {author} {\bibfnamefont {T.}~\bibnamefont
  {Byrnes}}, \bibinfo {author} {\bibfnamefont {N.~Y.}\ \bibnamefont {Kim}},\
  and\ \bibinfo {author} {\bibfnamefont {Y.}~\bibnamefont {Yamamoto}},\
  }\bibfield  {title} {\bibinfo {title} {Exciton–polariton condensates},\
  }\href {https://doi.org/10.1038/nphys3143} {\bibfield  {journal} {\bibinfo
  {journal} {Nat. Phys}\ }\textbf {\bibinfo {volume} {10}},\ \bibinfo {pages}
  {803–813} (\bibinfo {year} {2014})}\BibitemShut {NoStop}%
\bibitem [{\citenamefont {Mivehvar}\ \emph {et~al.}(2021)\citenamefont
  {Mivehvar}, \citenamefont {Piazza}, \citenamefont {Donner},\ and\
  \citenamefont {Ritsch}}]{Mivehvar2021}%
  \BibitemOpen
  \bibfield  {author} {\bibinfo {author} {\bibfnamefont {F.}~\bibnamefont
  {Mivehvar}}, \bibinfo {author} {\bibfnamefont {F.}~\bibnamefont {Piazza}},
  \bibinfo {author} {\bibfnamefont {T.}~\bibnamefont {Donner}},\ and\ \bibinfo
  {author} {\bibfnamefont {H.}~\bibnamefont {Ritsch}},\ }\bibfield  {title}
  {\bibinfo {title} {Cavity {QED} with quantum gases: new paradigms in
  many-body physics},\ }\href {https://doi.org/10.1080/00018732.2021.1969727}
  {\bibfield  {journal} {\bibinfo  {journal} {Adv. Phys.}\ }\textbf {\bibinfo
  {volume} {70}},\ \bibinfo {pages} {1} (\bibinfo {year} {2021})}\BibitemShut
  {NoStop}%
\bibitem [{\citenamefont {Carusotto}\ and\ \citenamefont
  {Ciuti}(2013)}]{Carusotto2013Quantum}%
  \BibitemOpen
  \bibfield  {author} {\bibinfo {author} {\bibfnamefont {I.}~\bibnamefont
  {Carusotto}}\ and\ \bibinfo {author} {\bibfnamefont {C.}~\bibnamefont
  {Ciuti}},\ }\bibfield  {title} {\bibinfo {title} {Quantum fluids of light},\
  }\href {https://doi.org/10.1103/RevModPhys.85.299} {\bibfield  {journal}
  {\bibinfo  {journal} {Rev. Mod. Phys.}\ }\textbf {\bibinfo {volume} {85}},\
  \bibinfo {pages} {299} (\bibinfo {year} {2013})}\BibitemShut {NoStop}%
\bibitem [{\citenamefont {Deng}\ \emph {et~al.}(2002)\citenamefont {Deng},
  \citenamefont {Weihs}, \citenamefont {Santori}, \citenamefont {Bloch},\ and\
  \citenamefont {Yamamoto}}]{Deng2002}%
  \BibitemOpen
  \bibfield  {author} {\bibinfo {author} {\bibfnamefont {H.}~\bibnamefont
  {Deng}}, \bibinfo {author} {\bibfnamefont {G.}~\bibnamefont {Weihs}},
  \bibinfo {author} {\bibfnamefont {C.}~\bibnamefont {Santori}}, \bibinfo
  {author} {\bibfnamefont {J.}~\bibnamefont {Bloch}},\ and\ \bibinfo {author}
  {\bibfnamefont {Y.}~\bibnamefont {Yamamoto}},\ }\bibfield  {title} {\bibinfo
  {title} {Condensation of semiconductor microcavity exciton polaritons},\
  }\href {https://doi.org/10.1126/science.1074464} {\bibfield  {journal}
  {\bibinfo  {journal} {Science}\ }\textbf {\bibinfo {volume} {298}},\ \bibinfo
  {pages} {199–202} (\bibinfo {year} {2002})}\BibitemShut {NoStop}%
\bibitem [{\citenamefont {Kasprzak}\ \emph {et~al.}(2006)\citenamefont
  {Kasprzak}, \citenamefont {Richard}, \citenamefont {Kundermann},
  \citenamefont {Baas}, \citenamefont {Jeambrun}, \citenamefont {Keeling},
  \citenamefont {Marchetti}, \citenamefont {Szymańska}, \citenamefont
  {André}, \citenamefont {Staehli}, \citenamefont {Savona}, \citenamefont
  {Littlewood}, \citenamefont {Deveaud},\ and\ \citenamefont
  {Dang}}]{Kasprzak2006}%
  \BibitemOpen
  \bibfield  {author} {\bibinfo {author} {\bibfnamefont {J.}~\bibnamefont
  {Kasprzak}}, \bibinfo {author} {\bibfnamefont {M.}~\bibnamefont {Richard}},
  \bibinfo {author} {\bibfnamefont {S.}~\bibnamefont {Kundermann}}, \bibinfo
  {author} {\bibfnamefont {A.}~\bibnamefont {Baas}}, \bibinfo {author}
  {\bibfnamefont {P.}~\bibnamefont {Jeambrun}}, \bibinfo {author}
  {\bibfnamefont {J.~M.~J.}\ \bibnamefont {Keeling}}, \bibinfo {author}
  {\bibfnamefont {F.~M.}\ \bibnamefont {Marchetti}}, \bibinfo {author}
  {\bibfnamefont {M.~H.}\ \bibnamefont {Szymańska}}, \bibinfo {author}
  {\bibfnamefont {R.}~\bibnamefont {André}}, \bibinfo {author} {\bibfnamefont
  {J.~L.}\ \bibnamefont {Staehli}}, \bibinfo {author} {\bibfnamefont
  {V.}~\bibnamefont {Savona}}, \bibinfo {author} {\bibfnamefont {P.~B.}\
  \bibnamefont {Littlewood}}, \bibinfo {author} {\bibfnamefont
  {B.}~\bibnamefont {Deveaud}},\ and\ \bibinfo {author} {\bibfnamefont {L.~S.}\
  \bibnamefont {Dang}},\ }\bibfield  {title} {\bibinfo {title}
  {Bose–{E}instein condensation of exciton polaritons},\ }\href
  {https://doi.org/10.1038/nature05131} {\bibfield  {journal} {\bibinfo
  {journal} {Nature}\ }\textbf {\bibinfo {volume} {443}},\ \bibinfo {pages}
  {409–414} (\bibinfo {year} {2006})}\BibitemShut {NoStop}%
\bibitem [{\citenamefont {Balili}\ \emph {et~al.}(2007)\citenamefont {Balili},
  \citenamefont {Hartwell}, \citenamefont {Snoke}, \citenamefont {Pfeiffer},\
  and\ \citenamefont {West}}]{Balili2007}%
  \BibitemOpen
  \bibfield  {author} {\bibinfo {author} {\bibfnamefont {R.}~\bibnamefont
  {Balili}}, \bibinfo {author} {\bibfnamefont {V.}~\bibnamefont {Hartwell}},
  \bibinfo {author} {\bibfnamefont {D.}~\bibnamefont {Snoke}}, \bibinfo
  {author} {\bibfnamefont {L.}~\bibnamefont {Pfeiffer}},\ and\ \bibinfo
  {author} {\bibfnamefont {K.}~\bibnamefont {West}},\ }\bibfield  {title}
  {\bibinfo {title} {Bose-{E}instein {C}ondensation of {M}icrocavity
  {P}olaritons in a {T}rap},\ }\href {https://doi.org/10.1126/science.1140990}
  {\bibfield  {journal} {\bibinfo  {journal} {Science}\ }\textbf {\bibinfo
  {volume} {316}},\ \bibinfo {pages} {1007–1010} (\bibinfo {year}
  {2007})}\BibitemShut {NoStop}%
\bibitem [{\citenamefont {Amo}\ \emph {et~al.}(2009)\citenamefont {Amo},
  \citenamefont {Lefrère}, \citenamefont {Pigeon}, \citenamefont {Adrados},
  \citenamefont {Ciuti}, \citenamefont {Carusotto}, \citenamefont {Houdré},
  \citenamefont {Giacobino},\ and\ \citenamefont {Bramati}}]{Amo2009}%
  \BibitemOpen
  \bibfield  {author} {\bibinfo {author} {\bibfnamefont {A.}~\bibnamefont
  {Amo}}, \bibinfo {author} {\bibfnamefont {J.}~\bibnamefont {Lefrère}},
  \bibinfo {author} {\bibfnamefont {S.}~\bibnamefont {Pigeon}}, \bibinfo
  {author} {\bibfnamefont {C.}~\bibnamefont {Adrados}}, \bibinfo {author}
  {\bibfnamefont {C.}~\bibnamefont {Ciuti}}, \bibinfo {author} {\bibfnamefont
  {I.}~\bibnamefont {Carusotto}}, \bibinfo {author} {\bibfnamefont
  {R.}~\bibnamefont {Houdré}}, \bibinfo {author} {\bibfnamefont
  {E.}~\bibnamefont {Giacobino}},\ and\ \bibinfo {author} {\bibfnamefont
  {A.}~\bibnamefont {Bramati}},\ }\bibfield  {title} {\bibinfo {title}
  {Superfluidity of polaritons in semiconductor microcavities},\ }\href
  {https://doi.org/10.1038/nphys1364} {\bibfield  {journal} {\bibinfo
  {journal} {Nat. Phys}\ }\textbf {\bibinfo {volume} {5}},\ \bibinfo {pages}
  {805–810} (\bibinfo {year} {2009})}\BibitemShut {NoStop}%
\bibitem [{\citenamefont {Klaers}\ \emph {et~al.}(2010)\citenamefont {Klaers},
  \citenamefont {Schmitt}, \citenamefont {Vewinger},\ and\ \citenamefont
  {Weitz}}]{Klaers2010}%
  \BibitemOpen
  \bibfield  {author} {\bibinfo {author} {\bibfnamefont {J.}~\bibnamefont
  {Klaers}}, \bibinfo {author} {\bibfnamefont {J.}~\bibnamefont {Schmitt}},
  \bibinfo {author} {\bibfnamefont {F.}~\bibnamefont {Vewinger}},\ and\
  \bibinfo {author} {\bibfnamefont {M.}~\bibnamefont {Weitz}},\ }\bibfield
  {title} {\bibinfo {title} {Bose–{E}instein condensation of photons in an
  optical microcavity},\ }\href {https://doi.org/10.1038/nature09567}
  {\bibfield  {journal} {\bibinfo  {journal} {Nature}\ }\textbf {\bibinfo
  {volume} {468}},\ \bibinfo {pages} {545–548} (\bibinfo {year}
  {2010})}\BibitemShut {NoStop}%
\bibitem [{\citenamefont {Hartmann}\ \emph {et~al.}(2008)\citenamefont
  {Hartmann}, \citenamefont {Brandão},\ and\ \citenamefont
  {Plenio}}]{Hartmann2008}%
  \BibitemOpen
  \bibfield  {author} {\bibinfo {author} {\bibfnamefont {M.}~\bibnamefont
  {Hartmann}}, \bibinfo {author} {\bibfnamefont {F.}~\bibnamefont {Brandão}},\
  and\ \bibinfo {author} {\bibfnamefont {M.}~\bibnamefont {Plenio}},\
  }\bibfield  {title} {\bibinfo {title} {Quantum many‐body phenomena in
  coupled cavity arrays},\ }\href {https://doi.org/10.1002/lpor.200810046}
  {\bibfield  {journal} {\bibinfo  {journal} {Laser \& Photon. Rev.}\ }\textbf
  {\bibinfo {volume} {2}},\ \bibinfo {pages} {527–556} (\bibinfo {year}
  {2008})}\BibitemShut {NoStop}%
\bibitem [{\citenamefont {Schmidt}\ and\ \citenamefont
  {Blatter}(2009)}]{Schmidt2009}%
  \BibitemOpen
  \bibfield  {author} {\bibinfo {author} {\bibfnamefont {S.}~\bibnamefont
  {Schmidt}}\ and\ \bibinfo {author} {\bibfnamefont {G.}~\bibnamefont
  {Blatter}},\ }\bibfield  {title} {\bibinfo {title} {Strong {C}oupling
  {T}heory for the {J}aynes-{C}ummings-{H}ubbard {M}odel},\ }\href
  {https://doi.org/10.1103/PhysRevLett.103.086403} {\bibfield  {journal}
  {\bibinfo  {journal} {Phys. Rev. Lett.}\ }\textbf {\bibinfo {volume} {103}},\
  \bibinfo {pages} {086403} (\bibinfo {year} {2009})}\BibitemShut {NoStop}%
\bibitem [{\citenamefont {Tomadin}\ \emph {et~al.}(2010)\citenamefont
  {Tomadin}, \citenamefont {Giovannetti}, \citenamefont {Fazio}, \citenamefont
  {Gerace}, \citenamefont {Carusotto}, \citenamefont {T\"ureci},\ and\
  \citenamefont {Imamoglu}}]{Tomadin2010}%
  \BibitemOpen
  \bibfield  {author} {\bibinfo {author} {\bibfnamefont {A.}~\bibnamefont
  {Tomadin}}, \bibinfo {author} {\bibfnamefont {V.}~\bibnamefont
  {Giovannetti}}, \bibinfo {author} {\bibfnamefont {R.}~\bibnamefont {Fazio}},
  \bibinfo {author} {\bibfnamefont {D.}~\bibnamefont {Gerace}}, \bibinfo
  {author} {\bibfnamefont {I.}~\bibnamefont {Carusotto}}, \bibinfo {author}
  {\bibfnamefont {H.~E.}\ \bibnamefont {T\"ureci}},\ and\ \bibinfo {author}
  {\bibfnamefont {A.}~\bibnamefont {Imamoglu}},\ }\bibfield  {title} {\bibinfo
  {title} {Signatures of the superfluid-insulator phase transition in
  laser-driven dissipative nonlinear cavity arrays},\ }\href
  {https://doi.org/10.1103/PhysRevA.81.061801} {\bibfield  {journal} {\bibinfo
  {journal} {Phys. Rev. A}\ }\textbf {\bibinfo {volume} {81}},\ \bibinfo
  {pages} {061801} (\bibinfo {year} {2010})}\BibitemShut {NoStop}%
\bibitem [{\citenamefont {Wu}\ \emph {et~al.}(2011)\citenamefont {Wu},
  \citenamefont {Gao}, \citenamefont {Deng}, \citenamefont {Dai}, \citenamefont
  {Chen},\ and\ \citenamefont {Li}}]{Wu2011}%
  \BibitemOpen
  \bibfield  {author} {\bibinfo {author} {\bibfnamefont {C.-W.}\ \bibnamefont
  {Wu}}, \bibinfo {author} {\bibfnamefont {M.}~\bibnamefont {Gao}}, \bibinfo
  {author} {\bibfnamefont {Z.-J.}\ \bibnamefont {Deng}}, \bibinfo {author}
  {\bibfnamefont {H.-Y.}\ \bibnamefont {Dai}}, \bibinfo {author} {\bibfnamefont
  {P.-X.}\ \bibnamefont {Chen}},\ and\ \bibinfo {author} {\bibfnamefont
  {C.-Z.}\ \bibnamefont {Li}},\ }\bibfield  {title} {\bibinfo {title} {Quantum
  phase transition of light in a one-dimensional photon-hopping-controllable
  resonator array},\ }\href {https://doi.org/10.1103/PhysRevA.84.043827}
  {\bibfield  {journal} {\bibinfo  {journal} {Phys. Rev. A}\ }\textbf {\bibinfo
  {volume} {84}},\ \bibinfo {pages} {043827} (\bibinfo {year}
  {2011})}\BibitemShut {NoStop}%
\bibitem [{foo({\natexlab{a}})}]{footnote1}%
  \BibitemOpen
  \href@noop {} {}\bibinfo {note} {Although these models share many
  similarities, such as bosons hopping coherently between nearest-neighbor
  sites with local nonlinearities, we note that quantitative differences may
  arise on different energy scales.}\BibitemShut {Stop}%
\bibitem [{\citenamefont {Le~Boit\'e}\ \emph {et~al.}(2013)\citenamefont
  {Le~Boit\'e}, \citenamefont {Orso},\ and\ \citenamefont
  {Ciuti}}]{LeBoit2013}%
  \BibitemOpen
  \bibfield  {author} {\bibinfo {author} {\bibfnamefont {A.}~\bibnamefont
  {Le~Boit\'e}}, \bibinfo {author} {\bibfnamefont {G.}~\bibnamefont {Orso}},\
  and\ \bibinfo {author} {\bibfnamefont {C.}~\bibnamefont {Ciuti}},\ }\bibfield
   {title} {\bibinfo {title} {Steady-{S}tate {P}hases and {T}unneling-{I}nduced
  {I}nstabilities in the {D}riven {D}issipative {B}ose-{H}ubbard {M}odel},\
  }\href {https://doi.org/10.1103/PhysRevLett.110.233601} {\bibfield  {journal}
  {\bibinfo  {journal} {Phys. Rev. Lett.}\ }\textbf {\bibinfo {volume} {110}},\
  \bibinfo {pages} {233601} (\bibinfo {year} {2013})}\BibitemShut {NoStop}%
\bibitem [{\citenamefont {Le~Boit\'e}\ \emph {et~al.}(2014)\citenamefont
  {Le~Boit\'e}, \citenamefont {Orso},\ and\ \citenamefont
  {Ciuti}}]{LeBoit2014}%
  \BibitemOpen
  \bibfield  {author} {\bibinfo {author} {\bibfnamefont {A.}~\bibnamefont
  {Le~Boit\'e}}, \bibinfo {author} {\bibfnamefont {G.}~\bibnamefont {Orso}},\
  and\ \bibinfo {author} {\bibfnamefont {C.}~\bibnamefont {Ciuti}},\ }\bibfield
   {title} {\bibinfo {title} {Bose-{H}ubbard model: {R}elation between
  driven-dissipative steady states and equilibrium quantum phases},\ }\href
  {https://doi.org/10.1103/PhysRevA.90.063821} {\bibfield  {journal} {\bibinfo
  {journal} {Phys. Rev. A}\ }\textbf {\bibinfo {volume} {90}},\ \bibinfo
  {pages} {063821} (\bibinfo {year} {2014})}\BibitemShut {NoStop}%
\bibitem [{\citenamefont {Gra\ss{}}(2019)}]{Grass2019}%
  \BibitemOpen
  \bibfield  {author} {\bibinfo {author} {\bibfnamefont {T.}~\bibnamefont
  {Gra\ss{}}},\ }\bibfield  {title} {\bibinfo {title} {Excitations and
  correlations in the driven-dissipative {B}ose-{H}ubbard model},\ }\href
  {https://doi.org/10.1103/PhysRevA.99.043607} {\bibfield  {journal} {\bibinfo
  {journal} {Phys. Rev. A}\ }\textbf {\bibinfo {volume} {99}},\ \bibinfo
  {pages} {043607} (\bibinfo {year} {2019})}\BibitemShut {NoStop}%
\bibitem [{\citenamefont {Lu}\ \emph {et~al.}(2014)\citenamefont {Lu},
  \citenamefont {Joannopoulos},\ and\ \citenamefont {Soljačić}}]{Lu2014}%
  \BibitemOpen
  \bibfield  {author} {\bibinfo {author} {\bibfnamefont {L.}~\bibnamefont
  {Lu}}, \bibinfo {author} {\bibfnamefont {J.~D.}\ \bibnamefont
  {Joannopoulos}},\ and\ \bibinfo {author} {\bibfnamefont {M.}~\bibnamefont
  {Soljačić}},\ }\bibfield  {title} {\bibinfo {title} {Topological
  photonics},\ }\href {https://doi.org/10.1038/nphoton.2014.248} {\bibfield
  {journal} {\bibinfo  {journal} {Nat. Photon}\ }\textbf {\bibinfo {volume}
  {8}},\ \bibinfo {pages} {821–829} (\bibinfo {year} {2014})}\BibitemShut
  {NoStop}%
\bibitem [{\citenamefont {Ozawa}\ \emph {et~al.}(2019)\citenamefont {Ozawa},
  \citenamefont {Price}, \citenamefont {Amo}, \citenamefont {Goldman},
  \citenamefont {Hafezi}, \citenamefont {Lu}, \citenamefont {Rechtsman},
  \citenamefont {Schuster}, \citenamefont {Simon}, \citenamefont {Zilberberg},\
  and\ \citenamefont {Carusotto}}]{Ozawa2019}%
  \BibitemOpen
  \bibfield  {author} {\bibinfo {author} {\bibfnamefont {T.}~\bibnamefont
  {Ozawa}}, \bibinfo {author} {\bibfnamefont {H.~M.}\ \bibnamefont {Price}},
  \bibinfo {author} {\bibfnamefont {A.}~\bibnamefont {Amo}}, \bibinfo {author}
  {\bibfnamefont {N.}~\bibnamefont {Goldman}}, \bibinfo {author} {\bibfnamefont
  {M.}~\bibnamefont {Hafezi}}, \bibinfo {author} {\bibfnamefont
  {L.}~\bibnamefont {Lu}}, \bibinfo {author} {\bibfnamefont {M.~C.}\
  \bibnamefont {Rechtsman}}, \bibinfo {author} {\bibfnamefont {D.}~\bibnamefont
  {Schuster}}, \bibinfo {author} {\bibfnamefont {J.}~\bibnamefont {Simon}},
  \bibinfo {author} {\bibfnamefont {O.}~\bibnamefont {Zilberberg}},\ and\
  \bibinfo {author} {\bibfnamefont {I.}~\bibnamefont {Carusotto}},\ }\bibfield
  {title} {\bibinfo {title} {Topological photonics},\ }\href
  {https://doi.org/10.1103/RevModPhys.91.015006} {\bibfield  {journal}
  {\bibinfo  {journal} {Rev. Mod. Phys.}\ }\textbf {\bibinfo {volume} {91}},\
  \bibinfo {pages} {015006} (\bibinfo {year} {2019})}\BibitemShut {NoStop}%
\bibitem [{\citenamefont {Wannier}(1960)}]{Wannier1960}%
  \BibitemOpen
  \bibfield  {author} {\bibinfo {author} {\bibfnamefont {G.~H.}\ \bibnamefont
  {Wannier}},\ }\bibfield  {title} {\bibinfo {title} {Wave {F}unctions and
  {E}ffective {H}amiltonian for {B}loch {E}lectrons in an {E}lectric {F}ield},\
  }\href {https://doi.org/10.1103/PhysRev.117.432} {\bibfield  {journal}
  {\bibinfo  {journal} {Phys. Rev.}\ }\textbf {\bibinfo {volume} {117}},\
  \bibinfo {pages} {432} (\bibinfo {year} {1960})}\BibitemShut {NoStop}%
\bibitem [{\citenamefont {Hartmann}\ \emph {et~al.}(2004)\citenamefont
  {Hartmann}, \citenamefont {Keck}, \citenamefont {Korsch},\ and\ \citenamefont
  {Mossmann}}]{Hartmann2004}%
  \BibitemOpen
  \bibfield  {author} {\bibinfo {author} {\bibfnamefont {T.}~\bibnamefont
  {Hartmann}}, \bibinfo {author} {\bibfnamefont {F.}~\bibnamefont {Keck}},
  \bibinfo {author} {\bibfnamefont {H.~J.}\ \bibnamefont {Korsch}},\ and\
  \bibinfo {author} {\bibfnamefont {S.}~\bibnamefont {Mossmann}},\ }\bibfield
  {title} {\bibinfo {title} {Dynamics of {B}loch oscillations},\ }\href
  {https://doi.org/10.1088/1367-2630/6/1/002} {\bibfield  {journal} {\bibinfo
  {journal} {New J. Phys.}\ }\textbf {\bibinfo {volume} {6}},\ \bibinfo {pages}
  {2–2} (\bibinfo {year} {2004})}\BibitemShut {NoStop}%
\bibitem [{\citenamefont {Peschel}\ \emph {et~al.}(1998)\citenamefont
  {Peschel}, \citenamefont {Pertsch},\ and\ \citenamefont
  {Lederer}}]{Peschel1998}%
  \BibitemOpen
  \bibfield  {author} {\bibinfo {author} {\bibfnamefont {U.}~\bibnamefont
  {Peschel}}, \bibinfo {author} {\bibfnamefont {T.}~\bibnamefont {Pertsch}},\
  and\ \bibinfo {author} {\bibfnamefont {F.}~\bibnamefont {Lederer}},\
  }\bibfield  {title} {\bibinfo {title} {Optical {B}loch oscillations in
  waveguide arrays},\ }\href {https://doi.org/10.1364/OL.23.001701} {\bibfield
  {journal} {\bibinfo  {journal} {Opt. Lett.}\ }\textbf {\bibinfo {volume}
  {23}},\ \bibinfo {pages} {1701} (\bibinfo {year} {1998})}\BibitemShut
  {NoStop}%
\bibitem [{\citenamefont {Sapienza}\ \emph {et~al.}(2003)\citenamefont
  {Sapienza}, \citenamefont {Costantino}, \citenamefont {Wiersma},
  \citenamefont {Ghulinyan}, \citenamefont {Oton},\ and\ \citenamefont
  {Pavesi}}]{Sapienza2003}%
  \BibitemOpen
  \bibfield  {author} {\bibinfo {author} {\bibfnamefont {R.}~\bibnamefont
  {Sapienza}}, \bibinfo {author} {\bibfnamefont {P.}~\bibnamefont
  {Costantino}}, \bibinfo {author} {\bibfnamefont {D.}~\bibnamefont {Wiersma}},
  \bibinfo {author} {\bibfnamefont {M.}~\bibnamefont {Ghulinyan}}, \bibinfo
  {author} {\bibfnamefont {C.~J.}\ \bibnamefont {Oton}},\ and\ \bibinfo
  {author} {\bibfnamefont {L.}~\bibnamefont {Pavesi}},\ }\bibfield  {title}
  {\bibinfo {title} {Optical {A}nalogue of {E}lectronic {B}loch
  {O}scillations},\ }\href {https://doi.org/10.1103/PhysRevLett.91.263902}
  {\bibfield  {journal} {\bibinfo  {journal} {Phys. Rev. Lett.}\ }\textbf
  {\bibinfo {volume} {91}},\ \bibinfo {pages} {263902} (\bibinfo {year}
  {2003})}\BibitemShut {NoStop}%
\bibitem [{\citenamefont {Yuan}\ and\ \citenamefont {Fan}(2016)}]{Yuan2016}%
  \BibitemOpen
  \bibfield  {author} {\bibinfo {author} {\bibfnamefont {L.}~\bibnamefont
  {Yuan}}\ and\ \bibinfo {author} {\bibfnamefont {S.}~\bibnamefont {Fan}},\
  }\bibfield  {title} {\bibinfo {title} {Bloch oscillation and unidirectional
  translation of frequency in a dynamically modulated ring resonator},\ }\href
  {https://doi.org/10.1364/OPTICA.3.001014} {\bibfield  {journal} {\bibinfo
  {journal} {Optica}\ }\textbf {\bibinfo {volume} {3}},\ \bibinfo {pages}
  {1014} (\bibinfo {year} {2016})}\BibitemShut {NoStop}%
\bibitem [{\citenamefont {Buchleitner}\ and\ \citenamefont
  {Kolovsky}(2003)}]{Buchleitner2003}%
  \BibitemOpen
  \bibfield  {author} {\bibinfo {author} {\bibfnamefont {A.}~\bibnamefont
  {Buchleitner}}\ and\ \bibinfo {author} {\bibfnamefont {A.~R.}\ \bibnamefont
  {Kolovsky}},\ }\bibfield  {title} {\bibinfo {title} {Interaction-{I}nduced
  {D}ecoherence of {A}tomic {B}loch {O}scillations},\ }\href
  {https://doi.org/10.1103/PhysRevLett.91.253002} {\bibfield  {journal}
  {\bibinfo  {journal} {Phys. Rev. Lett.}\ }\textbf {\bibinfo {volume} {91}},\
  \bibinfo {pages} {253002} (\bibinfo {year} {2003})}\BibitemShut {NoStop}%
\bibitem [{\citenamefont {Gustavsson}\ \emph {et~al.}(2008)\citenamefont
  {Gustavsson}, \citenamefont {Haller}, \citenamefont {Mark}, \citenamefont
  {Danzl}, \citenamefont {Rojas-Kopeinig},\ and\ \citenamefont
  {N\"agerl}}]{Gustavsson2008}%
  \BibitemOpen
  \bibfield  {author} {\bibinfo {author} {\bibfnamefont {M.}~\bibnamefont
  {Gustavsson}}, \bibinfo {author} {\bibfnamefont {E.}~\bibnamefont {Haller}},
  \bibinfo {author} {\bibfnamefont {M.~J.}\ \bibnamefont {Mark}}, \bibinfo
  {author} {\bibfnamefont {J.~G.}\ \bibnamefont {Danzl}}, \bibinfo {author}
  {\bibfnamefont {G.}~\bibnamefont {Rojas-Kopeinig}},\ and\ \bibinfo {author}
  {\bibfnamefont {H.-C.}\ \bibnamefont {N\"agerl}},\ }\bibfield  {title}
  {\bibinfo {title} {Control of {I}nteraction-{I}nduced {D}ephasing of {B}loch
  {O}scillations},\ }\href {https://doi.org/10.1103/PhysRevLett.100.080404}
  {\bibfield  {journal} {\bibinfo  {journal} {Phys. Rev. Lett.}\ }\textbf
  {\bibinfo {volume} {100}},\ \bibinfo {pages} {080404} (\bibinfo {year}
  {2008})}\BibitemShut {NoStop}%
\bibitem [{\citenamefont {Krimer}\ \emph {et~al.}(2009)\citenamefont {Krimer},
  \citenamefont {Khomeriki},\ and\ \citenamefont {Flach}}]{Krimer2009}%
  \BibitemOpen
  \bibfield  {author} {\bibinfo {author} {\bibfnamefont {D.~O.}\ \bibnamefont
  {Krimer}}, \bibinfo {author} {\bibfnamefont {R.}~\bibnamefont {Khomeriki}},\
  and\ \bibinfo {author} {\bibfnamefont {S.}~\bibnamefont {Flach}},\ }\bibfield
   {title} {\bibinfo {title} {Delocalization and spreading in a nonlinear
  {S}tark ladder},\ }\href {https://doi.org/10.1103/PhysRevE.80.036201}
  {\bibfield  {journal} {\bibinfo  {journal} {Phys. Rev. E}\ }\textbf {\bibinfo
  {volume} {80}},\ \bibinfo {pages} {036201} (\bibinfo {year}
  {2009})}\BibitemShut {NoStop}%
\bibitem [{\citenamefont {Eckstein}\ and\ \citenamefont
  {Werner}(2011)}]{Eckstein2011}%
  \BibitemOpen
  \bibfield  {author} {\bibinfo {author} {\bibfnamefont {M.}~\bibnamefont
  {Eckstein}}\ and\ \bibinfo {author} {\bibfnamefont {P.}~\bibnamefont
  {Werner}},\ }\bibfield  {title} {\bibinfo {title} {Damping of {B}loch
  {O}scillations in the {H}ubbard {M}odel},\ }\href
  {https://doi.org/10.1103/PhysRevLett.107.186406} {\bibfield  {journal}
  {\bibinfo  {journal} {Phys. Rev. Lett.}\ }\textbf {\bibinfo {volume} {107}},\
  \bibinfo {pages} {186406} (\bibinfo {year} {2011})}\BibitemShut {NoStop}%
\bibitem [{foo({\natexlab{b}})}]{footnote2}%
  \BibitemOpen
  \href@noop {} {}\bibinfo {note} {However, note that bound states of a few
  strongly interacting particles can experience fractional Bloch oscillations
  at frequencies that are multiples of single-particle Bloch oscillations
  \cite{Claro2003,Dias2007, Khomeriki2010,Wiater2017} (for experiments see
  \cite{Corrielli2013,Preiss2015}). Also, the decay of Blosch oscillations and
  the emergence of the directional motion in the excitation-phonon has been
  recently reported in \cite{Kosior2023}.}\BibitemShut {Stop}%
\bibitem [{\citenamefont {Houck}\ \emph {et~al.}(2012)\citenamefont {Houck},
  \citenamefont {T\"{u}reci},\ and\ \citenamefont {Koch}}]{Houck2012}%
  \BibitemOpen
  \bibfield  {author} {\bibinfo {author} {\bibfnamefont {A.~A.}\ \bibnamefont
  {Houck}}, \bibinfo {author} {\bibfnamefont {H.~E.}\ \bibnamefont
  {T\"{u}reci}},\ and\ \bibinfo {author} {\bibfnamefont {J.}~\bibnamefont
  {Koch}},\ }\bibfield  {title} {\bibinfo {title} {On-chip quantum simulation
  with superconducting circuits},\ }\href {https://doi.org/10.1038/nphys2251}
  {\bibfield  {journal} {\bibinfo  {journal} {Nat. Phys}\ }\textbf {\bibinfo
  {volume} {8}},\ \bibinfo {pages} {292–299} (\bibinfo {year}
  {2012})}\BibitemShut {NoStop}%
\bibitem [{\citenamefont {Jin}\ \emph {et~al.}(2013)\citenamefont {Jin},
  \citenamefont {Rossini}, \citenamefont {Fazio}, \citenamefont {Leib},\ and\
  \citenamefont {Hartmann}}]{Jin2013}%
  \BibitemOpen
  \bibfield  {author} {\bibinfo {author} {\bibfnamefont {J.}~\bibnamefont
  {Jin}}, \bibinfo {author} {\bibfnamefont {D.}~\bibnamefont {Rossini}},
  \bibinfo {author} {\bibfnamefont {R.}~\bibnamefont {Fazio}}, \bibinfo
  {author} {\bibfnamefont {M.}~\bibnamefont {Leib}},\ and\ \bibinfo {author}
  {\bibfnamefont {M.~J.}\ \bibnamefont {Hartmann}},\ }\bibfield  {title}
  {\bibinfo {title} {Photon {S}olid {P}hases in {D}riven {A}rrays of
  {N}onlinearly {C}oupled {C}avities},\ }\href
  {https://doi.org/10.1103/PhysRevLett.110.163605} {\bibfield  {journal}
  {\bibinfo  {journal} {Phys. Rev. Lett.}\ }\textbf {\bibinfo {volume} {110}},\
  \bibinfo {pages} {163605} (\bibinfo {year} {2013})}\BibitemShut {NoStop}%
\bibitem [{\citenamefont {Jin}\ \emph {et~al.}(2014)\citenamefont {Jin},
  \citenamefont {Rossini}, \citenamefont {Leib}, \citenamefont {Hartmann},\
  and\ \citenamefont {Fazio}}]{Jin2014}%
  \BibitemOpen
  \bibfield  {author} {\bibinfo {author} {\bibfnamefont {J.}~\bibnamefont
  {Jin}}, \bibinfo {author} {\bibfnamefont {D.}~\bibnamefont {Rossini}},
  \bibinfo {author} {\bibfnamefont {M.}~\bibnamefont {Leib}}, \bibinfo {author}
  {\bibfnamefont {M.~J.}\ \bibnamefont {Hartmann}},\ and\ \bibinfo {author}
  {\bibfnamefont {R.}~\bibnamefont {Fazio}},\ }\bibfield  {title} {\bibinfo
  {title} {Steady-state phase diagram of a driven {Q}{E}{D}-cavity array with
  cross-{K}err nonlinearities},\ }\href
  {https://doi.org/10.1103/PhysRevA.90.023827} {\bibfield  {journal} {\bibinfo
  {journal} {Phys. Rev. A}\ }\textbf {\bibinfo {volume} {90}},\ \bibinfo
  {pages} {023827} (\bibinfo {year} {2014})}\BibitemShut {NoStop}%
\bibitem [{\citenamefont {Majumdar}\ \emph {et~al.}(2012)\citenamefont
  {Majumdar}, \citenamefont {Rundquist}, \citenamefont {Bajcsy}, \citenamefont
  {Dasika}, \citenamefont {Bank},\ and\ \citenamefont {Vu\ifmmode
  \check{c}\else \v{c}\fi{}kovi\ifmmode~\acute{c}\else
  \'{c}\fi{}}}]{Majumdar2012}%
  \BibitemOpen
  \bibfield  {author} {\bibinfo {author} {\bibfnamefont {A.}~\bibnamefont
  {Majumdar}}, \bibinfo {author} {\bibfnamefont {A.}~\bibnamefont {Rundquist}},
  \bibinfo {author} {\bibfnamefont {M.}~\bibnamefont {Bajcsy}}, \bibinfo
  {author} {\bibfnamefont {V.~D.}\ \bibnamefont {Dasika}}, \bibinfo {author}
  {\bibfnamefont {S.~R.}\ \bibnamefont {Bank}},\ and\ \bibinfo {author}
  {\bibfnamefont {J.}~\bibnamefont {Vu\ifmmode \check{c}\else
  \v{c}\fi{}kovi\ifmmode~\acute{c}\else \'{c}\fi{}}},\ }\bibfield  {title}
  {\bibinfo {title} {Design and analysis of photonic crystal coupled cavity
  arrays for quantum simulation},\ }\href
  {https://doi.org/10.1103/PhysRevB.86.195312} {\bibfield  {journal} {\bibinfo
  {journal} {Phys. Rev. B}\ }\textbf {\bibinfo {volume} {86}},\ \bibinfo
  {pages} {195312} (\bibinfo {year} {2012})}\BibitemShut {NoStop}%
\bibitem [{\citenamefont {Lepert}\ \emph {et~al.}(2011)\citenamefont {Lepert},
  \citenamefont {Trupke}, \citenamefont {Hartmann}, \citenamefont {Plenio},\
  and\ \citenamefont {Hinds}}]{Lepert2011}%
  \BibitemOpen
  \bibfield  {author} {\bibinfo {author} {\bibfnamefont {G.}~\bibnamefont
  {Lepert}}, \bibinfo {author} {\bibfnamefont {M.}~\bibnamefont {Trupke}},
  \bibinfo {author} {\bibfnamefont {M.~J.}\ \bibnamefont {Hartmann}}, \bibinfo
  {author} {\bibfnamefont {M.~B.}\ \bibnamefont {Plenio}},\ and\ \bibinfo
  {author} {\bibfnamefont {E.~A.}\ \bibnamefont {Hinds}},\ }\bibfield  {title}
  {\bibinfo {title} {Arrays of waveguide-coupled optical cavities that interact
  strongly with atoms},\ }\href
  {https://doi.org/10.1088/1367-2630/13/11/113002} {\bibfield  {journal}
  {\bibinfo  {journal} {New J. Phys.}\ }\textbf {\bibinfo {volume} {13}},\
  \bibinfo {pages} {113002} (\bibinfo {year} {2011})}\BibitemShut {NoStop}%
\bibitem [{\citenamefont {Xiao}\ \emph {et~al.}(2008)\citenamefont {Xiao},
  \citenamefont {Gaddam},\ and\ \citenamefont {Yang}}]{Xiao2008}%
  \BibitemOpen
  \bibfield  {author} {\bibinfo {author} {\bibfnamefont {Y.-F.}\ \bibnamefont
  {Xiao}}, \bibinfo {author} {\bibfnamefont {V.}~\bibnamefont {Gaddam}},\ and\
  \bibinfo {author} {\bibfnamefont {L.}~\bibnamefont {Yang}},\ }\bibfield
  {title} {\bibinfo {title} {Coupled optical microcavities: an enhanced
  refractometric sensing configuration},\ }\href
  {https://doi.org/10.1364/oe.16.012538} {\bibfield  {journal} {\bibinfo
  {journal} {Optics Express}\ }\textbf {\bibinfo {volume} {16}},\ \bibinfo
  {pages} {12538} (\bibinfo {year} {2008})}\BibitemShut {NoStop}%
\bibitem [{\citenamefont {Rodriguez}\ \emph {et~al.}(2016)\citenamefont
  {Rodriguez}, \citenamefont {Amo}, \citenamefont {Sagnes}, \citenamefont
  {Le~Gratiet}, \citenamefont {Galopin}, \citenamefont {Lemaître},\ and\
  \citenamefont {Bloch}}]{Rodriguez2016}%
  \BibitemOpen
  \bibfield  {author} {\bibinfo {author} {\bibfnamefont {S.~R.~K.}\
  \bibnamefont {Rodriguez}}, \bibinfo {author} {\bibfnamefont {A.}~\bibnamefont
  {Amo}}, \bibinfo {author} {\bibfnamefont {I.}~\bibnamefont {Sagnes}},
  \bibinfo {author} {\bibfnamefont {L.}~\bibnamefont {Le~Gratiet}}, \bibinfo
  {author} {\bibfnamefont {E.}~\bibnamefont {Galopin}}, \bibinfo {author}
  {\bibfnamefont {A.}~\bibnamefont {Lemaître}},\ and\ \bibinfo {author}
  {\bibfnamefont {J.}~\bibnamefont {Bloch}},\ }\bibfield  {title} {\bibinfo
  {title} {Interaction-induced hopping phase in driven-dissipative coupled
  photonic microcavities},\ }\bibfield  {journal} {\bibinfo  {journal} {Nature
  Communications}\ }\textbf {\bibinfo {volume} {7}},\ \href
  {https://doi.org/10.1038/ncomms11887} {10.1038/ncomms11887} (\bibinfo {year}
  {2016})\BibitemShut {NoStop}%
\bibitem [{\citenamefont {Schneider}\ \emph {et~al.}(2016)\citenamefont
  {Schneider}, \citenamefont {Winkler}, \citenamefont {Fraser}, \citenamefont
  {Kamp}, \citenamefont {Yamamoto}, \citenamefont {Ostrovskaya},\ and\
  \citenamefont {H\"{o}fling}}]{Schneider2016}%
  \BibitemOpen
  \bibfield  {author} {\bibinfo {author} {\bibfnamefont {C.}~\bibnamefont
  {Schneider}}, \bibinfo {author} {\bibfnamefont {K.}~\bibnamefont {Winkler}},
  \bibinfo {author} {\bibfnamefont {M.~D.}\ \bibnamefont {Fraser}}, \bibinfo
  {author} {\bibfnamefont {M.}~\bibnamefont {Kamp}}, \bibinfo {author}
  {\bibfnamefont {Y.}~\bibnamefont {Yamamoto}}, \bibinfo {author}
  {\bibfnamefont {E.~A.}\ \bibnamefont {Ostrovskaya}},\ and\ \bibinfo {author}
  {\bibfnamefont {S.}~\bibnamefont {H\"{o}fling}},\ }\bibfield  {title}
  {\bibinfo {title} {Exciton-polariton trapping and potential landscape
  engineering},\ }\href {https://doi.org/10.1088/0034-4885/80/1/016503}
  {\bibfield  {journal} {\bibinfo  {journal} {Reports on Progress in Physics}\
  }\textbf {\bibinfo {volume} {80}},\ \bibinfo {pages} {016503} (\bibinfo
  {year} {2016})}\BibitemShut {NoStop}%
\bibitem [{\citenamefont {Douglas}\ \emph {et~al.}(2015)\citenamefont
  {Douglas}, \citenamefont {Habibian}, \citenamefont {Hung}, \citenamefont
  {Gorshkov}, \citenamefont {Kimble},\ and\ \citenamefont
  {Chang}}]{Douglas2015}%
  \BibitemOpen
  \bibfield  {author} {\bibinfo {author} {\bibfnamefont {J.~S.}\ \bibnamefont
  {Douglas}}, \bibinfo {author} {\bibfnamefont {H.}~\bibnamefont {Habibian}},
  \bibinfo {author} {\bibfnamefont {C.-L.}\ \bibnamefont {Hung}}, \bibinfo
  {author} {\bibfnamefont {A.~V.}\ \bibnamefont {Gorshkov}}, \bibinfo {author}
  {\bibfnamefont {H.~J.}\ \bibnamefont {Kimble}},\ and\ \bibinfo {author}
  {\bibfnamefont {D.~E.}\ \bibnamefont {Chang}},\ }\bibfield  {title} {\bibinfo
  {title} {Quantum many-body models with cold atoms coupled to photonic
  crystals},\ }\href {https://doi.org/10.1038/nphoton.2015.57} {\bibfield
  {journal} {\bibinfo  {journal} {Nature Photonics}\ }\textbf {\bibinfo
  {volume} {9}},\ \bibinfo {pages} {326–331} (\bibinfo {year}
  {2015})}\BibitemShut {NoStop}%
\bibitem [{\citenamefont {Kollár}\ \emph {et~al.}(2015)\citenamefont
  {Kollár}, \citenamefont {Papageorge}, \citenamefont {Baumann}, \citenamefont
  {Armen},\ and\ \citenamefont {Lev}}]{Kollr2015}%
  \BibitemOpen
  \bibfield  {author} {\bibinfo {author} {\bibfnamefont {A.~J.}\ \bibnamefont
  {Kollár}}, \bibinfo {author} {\bibfnamefont {A.~T.}\ \bibnamefont
  {Papageorge}}, \bibinfo {author} {\bibfnamefont {K.}~\bibnamefont {Baumann}},
  \bibinfo {author} {\bibfnamefont {M.~A.}\ \bibnamefont {Armen}},\ and\
  \bibinfo {author} {\bibfnamefont {B.~L.}\ \bibnamefont {Lev}},\ }\bibfield
  {title} {\bibinfo {title} {An adjustable-length cavity and bose–einstein
  condensate apparatus for multimode cavity qed},\ }\href
  {https://doi.org/10.1088/1367-2630/17/4/043012} {\bibfield  {journal}
  {\bibinfo  {journal} {New J. Phys.}\ }\textbf {\bibinfo {volume} {17}},\
  \bibinfo {pages} {043012} (\bibinfo {year} {2015})}\BibitemShut {NoStop}%
\bibitem [{\citenamefont {Zelan}\ \emph {et~al.}(2010)\citenamefont {Zelan},
  \citenamefont {Hagman}, \citenamefont {Karlsson}, \citenamefont {Dion},\ and\
  \citenamefont {Kastberg}}]{Zelan2010}%
  \BibitemOpen
  \bibfield  {author} {\bibinfo {author} {\bibfnamefont {M.}~\bibnamefont
  {Zelan}}, \bibinfo {author} {\bibfnamefont {H.}~\bibnamefont {Hagman}},
  \bibinfo {author} {\bibfnamefont {K.}~\bibnamefont {Karlsson}}, \bibinfo
  {author} {\bibfnamefont {C.~M.}\ \bibnamefont {Dion}},\ and\ \bibinfo
  {author} {\bibfnamefont {A.}~\bibnamefont {Kastberg}},\ }\bibfield  {title}
  {\bibinfo {title} {Fluctuation-induced drift in a gravitationally tilted
  optical lattice},\ }\href {https://doi.org/10.1103/PhysRevE.82.031136}
  {\bibfield  {journal} {\bibinfo  {journal} {Phys. Rev. E}\ }\textbf {\bibinfo
  {volume} {82}},\ \bibinfo {pages} {031136} (\bibinfo {year}
  {2010})}\BibitemShut {NoStop}%
\bibitem [{\citenamefont {Sharma}\ and\ \citenamefont
  {Mueller}(2021)}]{Sharma2021}%
  \BibitemOpen
  \bibfield  {author} {\bibinfo {author} {\bibfnamefont {V.}~\bibnamefont
  {Sharma}}\ and\ \bibinfo {author} {\bibfnamefont {E.~J.}\ \bibnamefont
  {Mueller}},\ }\bibfield  {title} {\bibinfo {title} {Driven-dissipative
  control of cold atoms in tilted optical lattices},\ }\href
  {https://doi.org/10.1103/PhysRevA.103.043322} {\bibfield  {journal} {\bibinfo
   {journal} {Phys. Rev. A}\ }\textbf {\bibinfo {volume} {103}},\ \bibinfo
  {pages} {043322} (\bibinfo {year} {2021})}\BibitemShut {NoStop}%
\bibitem [{SM()}]{SM}%
  \BibitemOpen
  \href@noop {} {}\bibinfo {note} {See Supplemental Materials for the
  derivation of the rotating frame Hamiltonian, discussion of Wannier-Stark
  states and the U(1) symmetry breaking terms, details of the cumulant
  expansion, and more examples of interaction-induced Bloch oscillations within
  and beyond the mean-field regime.}\BibitemShut {Stop}%
\bibitem [{\citenamefont {Yao}\ and\ \citenamefont
  {Zakrzewski}(2020)}]{Yao2020}%
  \BibitemOpen
  \bibfield  {author} {\bibinfo {author} {\bibfnamefont {R.}~\bibnamefont
  {Yao}}\ and\ \bibinfo {author} {\bibfnamefont {J.}~\bibnamefont
  {Zakrzewski}},\ }\bibfield  {title} {\bibinfo {title} {Many-body localization
  of bosons in an optical lattice: {D}ynamics in disorder-free potentials},\
  }\href {https://doi.org/10.1103/PhysRevB.102.104203} {\bibfield  {journal}
  {\bibinfo  {journal} {Phys. Rev. B}\ }\textbf {\bibinfo {volume} {102}},\
  \bibinfo {pages} {104203} (\bibinfo {year} {2020})}\BibitemShut {NoStop}%
\bibitem [{\citenamefont {Taylor}\ \emph {et~al.}(2020)\citenamefont {Taylor},
  \citenamefont {Schulz}, \citenamefont {Pollmann},\ and\ \citenamefont
  {Moessner}}]{Taylor2020}%
  \BibitemOpen
  \bibfield  {author} {\bibinfo {author} {\bibfnamefont {S.~R.}\ \bibnamefont
  {Taylor}}, \bibinfo {author} {\bibfnamefont {M.}~\bibnamefont {Schulz}},
  \bibinfo {author} {\bibfnamefont {F.}~\bibnamefont {Pollmann}},\ and\
  \bibinfo {author} {\bibfnamefont {R.}~\bibnamefont {Moessner}},\ }\bibfield
  {title} {\bibinfo {title} {Experimental probes of {S}tark many-body
  localization},\ }\href {https://doi.org/10.1103/PhysRevB.102.054206}
  {\bibfield  {journal} {\bibinfo  {journal} {Phys. Rev. B}\ }\textbf {\bibinfo
  {volume} {102}},\ \bibinfo {pages} {054206} (\bibinfo {year}
  {2020})}\BibitemShut {NoStop}%
\bibitem [{\citenamefont {Morong}\ \emph {et~al.}(2021)\citenamefont {Morong},
  \citenamefont {Liu}, \citenamefont {Becker}, \citenamefont {Collins},
  \citenamefont {Feng}, \citenamefont {Kyprianidis}, \citenamefont {Pagano},
  \citenamefont {You}, \citenamefont {Gorshkov},\ and\ \citenamefont
  {Monroe}}]{Morong2021}%
  \BibitemOpen
  \bibfield  {author} {\bibinfo {author} {\bibfnamefont {W.}~\bibnamefont
  {Morong}}, \bibinfo {author} {\bibfnamefont {F.}~\bibnamefont {Liu}},
  \bibinfo {author} {\bibfnamefont {P.}~\bibnamefont {Becker}}, \bibinfo
  {author} {\bibfnamefont {K.~S.}\ \bibnamefont {Collins}}, \bibinfo {author}
  {\bibfnamefont {L.}~\bibnamefont {Feng}}, \bibinfo {author} {\bibfnamefont
  {A.}~\bibnamefont {Kyprianidis}}, \bibinfo {author} {\bibfnamefont
  {G.}~\bibnamefont {Pagano}}, \bibinfo {author} {\bibfnamefont
  {T.}~\bibnamefont {You}}, \bibinfo {author} {\bibfnamefont {A.~V.}\
  \bibnamefont {Gorshkov}},\ and\ \bibinfo {author} {\bibfnamefont
  {C.}~\bibnamefont {Monroe}},\ }\bibfield  {title} {\bibinfo {title}
  {Observation of {S}tark many-body localization without disorder},\ }\href
  {https://doi.org/10.1038/s41586-021-03988-0} {\bibfield  {journal} {\bibinfo
  {journal} {Nature}\ }\textbf {\bibinfo {volume} {599}},\ \bibinfo {pages}
  {393–398} (\bibinfo {year} {2021})}\BibitemShut {NoStop}%
\bibitem [{\citenamefont {Plankensteiner}\ \emph {et~al.}(2022)\citenamefont
  {Plankensteiner}, \citenamefont {Hotter},\ and\ \citenamefont
  {Ritsch}}]{Plankensteiner2022}%
  \BibitemOpen
  \bibfield  {author} {\bibinfo {author} {\bibfnamefont {D.}~\bibnamefont
  {Plankensteiner}}, \bibinfo {author} {\bibfnamefont {C.}~\bibnamefont
  {Hotter}},\ and\ \bibinfo {author} {\bibfnamefont {H.}~\bibnamefont
  {Ritsch}},\ }\bibfield  {title} {\bibinfo {title} {Quantum{C}umulants.jl: {A}
  {J}ulia framework for generalized mean-field equations in open quantum
  systems},\ }\href {https://doi.org/10.22331/q-2022-01-04-617} {\bibfield
  {journal} {\bibinfo  {journal} {Quantum}\ }\textbf {\bibinfo {volume} {6}},\
  \bibinfo {pages} {617} (\bibinfo {year} {2022})}\BibitemShut {NoStop}%
\bibitem [{\citenamefont {Pethick}\ and\ \citenamefont
  {Smith}(2008)}]{Pethick2008}%
  \BibitemOpen
  \bibfield  {author} {\bibinfo {author} {\bibfnamefont {C.~J.}\ \bibnamefont
  {Pethick}}\ and\ \bibinfo {author} {\bibfnamefont {H.}~\bibnamefont
  {Smith}},\ }\href {https://doi.org/10.1017/cbo9780511802850} {\emph {\bibinfo
  {title} {Bose–Einstein Condensation in Dilute Gases}}}\ (\bibinfo
  {publisher} {Cambridge University Press},\ \bibinfo {year}
  {2008})\BibitemShut {NoStop}%
\bibitem [{\citenamefont {Polak}\ and\ \citenamefont
  {Kopeć}(2005)}]{Polak2005}%
  \BibitemOpen
  \bibfield  {author} {\bibinfo {author} {\bibfnamefont {T.~P.}\ \bibnamefont
  {Polak}}\ and\ \bibinfo {author} {\bibfnamefont {T.~K.}\ \bibnamefont
  {Kopeć}},\ }\bibfield  {title} {\bibinfo {title} {Local dissipation effects
  in two-dimensional quantum josephson junction arrays with a magnetic field},\
  }\bibfield  {journal} {\bibinfo  {journal} {Physical Review B}\ }\textbf
  {\bibinfo {volume} {72}},\ \href {https://doi.org/10.1103/physrevb.72.014509}
  {10.1103/physrevb.72.014509} (\bibinfo {year} {2005})\BibitemShut {NoStop}%
\bibitem [{\citenamefont {Huybrechts}\ \emph {et~al.}(2020)\citenamefont
  {Huybrechts}, \citenamefont {Minganti}, \citenamefont {Nori}, \citenamefont
  {Wouters},\ and\ \citenamefont {Shammah}}]{Huybrechts2020}%
  \BibitemOpen
  \bibfield  {author} {\bibinfo {author} {\bibfnamefont {D.}~\bibnamefont
  {Huybrechts}}, \bibinfo {author} {\bibfnamefont {F.}~\bibnamefont
  {Minganti}}, \bibinfo {author} {\bibfnamefont {F.}~\bibnamefont {Nori}},
  \bibinfo {author} {\bibfnamefont {M.}~\bibnamefont {Wouters}},\ and\ \bibinfo
  {author} {\bibfnamefont {N.}~\bibnamefont {Shammah}},\ }\bibfield  {title}
  {\bibinfo {title} {Validity of mean-field theory in a dissipative critical
  system: Liouvillian gap, $\mathbb{PT}$-symmetric antigap, and permutational
  symmetry in the $\mathit{XYZ}$ model},\ }\href
  {https://doi.org/10.1103/PhysRevB.101.214302} {\bibfield  {journal} {\bibinfo
   {journal} {Phys. Rev. B}\ }\textbf {\bibinfo {volume} {101}},\ \bibinfo
  {pages} {214302} (\bibinfo {year} {2020})}\BibitemShut {NoStop}%
\bibitem [{\citenamefont {Piazza}\ and\ \citenamefont
  {Ritsch}(2015)}]{Piazza2015}%
  \BibitemOpen
  \bibfield  {author} {\bibinfo {author} {\bibfnamefont {F.}~\bibnamefont
  {Piazza}}\ and\ \bibinfo {author} {\bibfnamefont {H.}~\bibnamefont
  {Ritsch}},\ }\bibfield  {title} {\bibinfo {title} {Self-{O}rdered {L}imit
  {C}ycles, {C}haos, and {P}hase {S}lippage with a {S}uperfluid inside an
  {O}ptical {R}esonator},\ }\href
  {https://doi.org/10.1103/PhysRevLett.115.163601} {\bibfield  {journal}
  {\bibinfo  {journal} {Phys. Rev. Lett.}\ }\textbf {\bibinfo {volume} {115}},\
  \bibinfo {pages} {163601} (\bibinfo {year} {2015})}\BibitemShut {NoStop}%
\bibitem [{\citenamefont {Chan}\ \emph {et~al.}(2015)\citenamefont {Chan},
  \citenamefont {Lee},\ and\ \citenamefont {Gopalakrishnan}}]{Chan2015}%
  \BibitemOpen
  \bibfield  {author} {\bibinfo {author} {\bibfnamefont {C.-K.}\ \bibnamefont
  {Chan}}, \bibinfo {author} {\bibfnamefont {T.~E.}\ \bibnamefont {Lee}},\ and\
  \bibinfo {author} {\bibfnamefont {S.}~\bibnamefont {Gopalakrishnan}},\
  }\bibfield  {title} {\bibinfo {title} {Limit-cycle phase in
  driven-dissipative spin systems},\ }\href
  {https://doi.org/10.1103/PhysRevA.91.051601} {\bibfield  {journal} {\bibinfo
  {journal} {Phys. Rev. A}\ }\textbf {\bibinfo {volume} {91}},\ \bibinfo
  {pages} {051601} (\bibinfo {year} {2015})}\BibitemShut {NoStop}%
\bibitem [{\citenamefont {Owen}\ \emph {et~al.}(2018)\citenamefont {Owen},
  \citenamefont {Jin}, \citenamefont {Rossini}, \citenamefont {Fazio},\ and\
  \citenamefont {Hartmann}}]{Owen2018}%
  \BibitemOpen
  \bibfield  {author} {\bibinfo {author} {\bibfnamefont {E.~T.}\ \bibnamefont
  {Owen}}, \bibinfo {author} {\bibfnamefont {J.}~\bibnamefont {Jin}}, \bibinfo
  {author} {\bibfnamefont {D.}~\bibnamefont {Rossini}}, \bibinfo {author}
  {\bibfnamefont {R.}~\bibnamefont {Fazio}},\ and\ \bibinfo {author}
  {\bibfnamefont {M.~J.}\ \bibnamefont {Hartmann}},\ }\bibfield  {title}
  {\bibinfo {title} {Quantum correlations and limit cycles in the
  driven-dissipative {H}eisenberg lattice},\ }\href
  {https://doi.org/10.1088/1367-2630/aab7d3} {\bibfield  {journal} {\bibinfo
  {journal} {New J. Phys.}\ }\textbf {\bibinfo {volume} {20}},\ \bibinfo
  {pages} {045004} (\bibinfo {year} {2018})}\BibitemShut {NoStop}%
\bibitem [{\citenamefont {Ke\ss{}ler}\ \emph {et~al.}(2019)\citenamefont
  {Ke\ss{}ler}, \citenamefont {Cosme}, \citenamefont {Hemmerling},
  \citenamefont {Mathey},\ and\ \citenamefont {Hemmerich}}]{Kessler2019}%
  \BibitemOpen
  \bibfield  {author} {\bibinfo {author} {\bibfnamefont {H.}~\bibnamefont
  {Ke\ss{}ler}}, \bibinfo {author} {\bibfnamefont {J.~G.}\ \bibnamefont
  {Cosme}}, \bibinfo {author} {\bibfnamefont {M.}~\bibnamefont {Hemmerling}},
  \bibinfo {author} {\bibfnamefont {L.}~\bibnamefont {Mathey}},\ and\ \bibinfo
  {author} {\bibfnamefont {A.}~\bibnamefont {Hemmerich}},\ }\bibfield  {title}
  {\bibinfo {title} {Emergent limit cycles and time crystal dynamics in an
  atom-cavity system},\ }\href {https://doi.org/10.1103/PhysRevA.99.053605}
  {\bibfield  {journal} {\bibinfo  {journal} {Phys. Rev. A}\ }\textbf {\bibinfo
  {volume} {99}},\ \bibinfo {pages} {053605} (\bibinfo {year}
  {2019})}\BibitemShut {NoStop}%
\bibitem [{\citenamefont {Colella}\ \emph {et~al.}(2022)\citenamefont
  {Colella}, \citenamefont {Kosior}, \citenamefont {Mivehvar},\ and\
  \citenamefont {Ritsch}}]{Colella2022}%
  \BibitemOpen
  \bibfield  {author} {\bibinfo {author} {\bibfnamefont {E.}~\bibnamefont
  {Colella}}, \bibinfo {author} {\bibfnamefont {A.}~\bibnamefont {Kosior}},
  \bibinfo {author} {\bibfnamefont {F.}~\bibnamefont {Mivehvar}},\ and\
  \bibinfo {author} {\bibfnamefont {H.}~\bibnamefont {Ritsch}},\ }\bibfield
  {title} {\bibinfo {title} {Open {Q}uantum {S}ystem {S}imulation of
  {F}araday’s {I}nduction {L}aw via {D}ynamical {I}nstabilities},\ }\href
  {https://doi.org/10.1103/PhysRevLett.128.070603} {\bibfield  {journal}
  {\bibinfo  {journal} {Phys. Rev. Lett.}\ }\textbf {\bibinfo {volume} {128}},\
  \bibinfo {pages} {070603} (\bibinfo {year} {2022})}\BibitemShut {NoStop}%
\bibitem [{\citenamefont {Gao}\ \emph {et~al.}(2023)\citenamefont {Gao},
  \citenamefont {Zhou}, \citenamefont {Guo},\ and\ \citenamefont
  {Luo}}]{Gao2023}%
  \BibitemOpen
  \bibfield  {author} {\bibinfo {author} {\bibfnamefont {P.}~\bibnamefont
  {Gao}}, \bibinfo {author} {\bibfnamefont {Z.-W.}\ \bibnamefont {Zhou}},
  \bibinfo {author} {\bibfnamefont {G.-C.}\ \bibnamefont {Guo}},\ and\ \bibinfo
  {author} {\bibfnamefont {X.-W.}\ \bibnamefont {Luo}},\ }\bibfield  {title}
  {\bibinfo {title} {Self-organized limit cycles in red-detuned atom-cavity
  systems},\ }\href {https://doi.org/10.1103/PhysRevA.107.023311} {\bibfield
  {journal} {\bibinfo  {journal} {Phys. Rev. A}\ }\textbf {\bibinfo {volume}
  {107}},\ \bibinfo {pages} {023311} (\bibinfo {year} {2023})}\BibitemShut
  {NoStop}%
\bibitem [{\citenamefont {Kosior}\ \emph
  {et~al.}(2023{\natexlab{a}})\citenamefont {Kosior}, \citenamefont {Ritsch},\
  and\ \citenamefont {Mivehvar}}]{Kosior2023b}%
  \BibitemOpen
  \bibfield  {author} {\bibinfo {author} {\bibfnamefont {A.}~\bibnamefont
  {Kosior}}, \bibinfo {author} {\bibfnamefont {H.}~\bibnamefont {Ritsch}},\
  and\ \bibinfo {author} {\bibfnamefont {F.}~\bibnamefont {Mivehvar}},\
  }\bibfield  {title} {\bibinfo {title} {Nonequilibrium phases of ultracold
  bosons with cavity-induced dynamic gauge fields},\ }\href
  {https://doi.org/10.21468/scipostphys.15.2.046} {\bibfield  {journal}
  {\bibinfo  {journal} {Sci{P}ost {P}hys.}\ }\textbf {\bibinfo {volume} {15}},\
  \bibinfo {pages} {046} (\bibinfo {year} {2023}{\natexlab{a}})}\BibitemShut
  {NoStop}%
\bibitem [{\citenamefont {Mivehvar}(2024)}]{Mivehvar2024}%
  \BibitemOpen
  \bibfield  {author} {\bibinfo {author} {\bibfnamefont {F.}~\bibnamefont
  {Mivehvar}},\ }\bibfield  {title} {\bibinfo {title} {Conventional and
  {U}nconventional {D}icke {M}odels: {M}ultistabilities and {N}onequilibrium
  {D}ynamics},\ }\href {https://doi.org/10.1103/PhysRevLett.132.073602}
  {\bibfield  {journal} {\bibinfo  {journal} {Phys. Rev. Lett.}\ }\textbf
  {\bibinfo {volume} {132}},\ \bibinfo {pages} {073602} (\bibinfo {year}
  {2024})}\BibitemShut {NoStop}%
\bibitem [{\citenamefont {Skulte}\ \emph {et~al.}(2024)\citenamefont {Skulte},
  \citenamefont {Kongkhambut}, \citenamefont {Keßler}, \citenamefont
  {Hemmerich}, \citenamefont {Mathey},\ and\ \citenamefont
  {Cosme}}]{Skulte2024}%
  \BibitemOpen
  \bibfield  {author} {\bibinfo {author} {\bibfnamefont {J.}~\bibnamefont
  {Skulte}}, \bibinfo {author} {\bibfnamefont {P.}~\bibnamefont {Kongkhambut}},
  \bibinfo {author} {\bibfnamefont {H.}~\bibnamefont {Keßler}}, \bibinfo
  {author} {\bibfnamefont {A.}~\bibnamefont {Hemmerich}}, \bibinfo {author}
  {\bibfnamefont {L.}~\bibnamefont {Mathey}},\ and\ \bibinfo {author}
  {\bibfnamefont {J.~G.}\ \bibnamefont {Cosme}},\ }\href@noop {} {\bibinfo
  {title} {Realizing limit cycles in dissipative bosonic systems}} (\bibinfo
  {year} {2024}),\ \Eprint {https://arxiv.org/abs/2401.05332} {arXiv:2401.05332
  [cond-mat.quant-gas]} \BibitemShut {NoStop}%
\bibitem [{\citenamefont {Phillips}\ \emph {et~al.}(2001)\citenamefont
  {Phillips}, \citenamefont {Fleischhauer}, \citenamefont {Mair}, \citenamefont
  {Walsworth},\ and\ \citenamefont {Lukin}}]{storage2001atomicvapor}%
  \BibitemOpen
  \bibfield  {author} {\bibinfo {author} {\bibfnamefont {D.~F.}\ \bibnamefont
  {Phillips}}, \bibinfo {author} {\bibfnamefont {A.}~\bibnamefont
  {Fleischhauer}}, \bibinfo {author} {\bibfnamefont {A.}~\bibnamefont {Mair}},
  \bibinfo {author} {\bibfnamefont {R.~L.}\ \bibnamefont {Walsworth}},\ and\
  \bibinfo {author} {\bibfnamefont {M.~D.}\ \bibnamefont {Lukin}},\ }\bibfield
  {title} {\bibinfo {title} {Storage of {L}ight in {A}tomic {V}apor},\ }\href
  {https://doi.org/10.1103/PhysRevLett.86.783} {\bibfield  {journal} {\bibinfo
  {journal} {Phys. Rev. Lett.}\ }\textbf {\bibinfo {volume} {86}},\ \bibinfo
  {pages} {783} (\bibinfo {year} {2001})}\BibitemShut {NoStop}%
\bibitem [{\citenamefont {Dudin}\ \emph {et~al.}(2013)\citenamefont {Dudin},
  \citenamefont {Li},\ and\ \citenamefont {Kuzmich}}]{lightstorage2013minute}%
  \BibitemOpen
  \bibfield  {author} {\bibinfo {author} {\bibfnamefont {Y.~O.}\ \bibnamefont
  {Dudin}}, \bibinfo {author} {\bibfnamefont {L.}~\bibnamefont {Li}},\ and\
  \bibinfo {author} {\bibfnamefont {A.}~\bibnamefont {Kuzmich}},\ }\bibfield
  {title} {\bibinfo {title} {Light storage on the time scale of a minute},\
  }\href {https://doi.org/10.1103/PhysRevA.87.031801} {\bibfield  {journal}
  {\bibinfo  {journal} {Phys. Rev. A}\ }\textbf {\bibinfo {volume} {87}},\
  \bibinfo {pages} {031801} (\bibinfo {year} {2013})}\BibitemShut {NoStop}%
\bibitem [{\citenamefont {Katz}\ and\ \citenamefont
  {Firstenberg}(2018)}]{lightstorage2018vapor}%
  \BibitemOpen
  \bibfield  {author} {\bibinfo {author} {\bibfnamefont {O.}~\bibnamefont
  {Katz}}\ and\ \bibinfo {author} {\bibfnamefont {O.}~\bibnamefont
  {Firstenberg}},\ }\bibfield  {title} {\bibinfo {title} {Light storage for one
  second in room-temperature alkali vapor},\ }\href
  {https://doi.org/10.1038/s41467-018-04458-4} {\bibfield  {journal} {\bibinfo
  {journal} {Nat. Commun}\ }\textbf {\bibinfo {volume} {9}},\ \bibinfo {pages}
  {2074} (\bibinfo {year} {2018})}\BibitemShut {NoStop}%
\bibitem [{\citenamefont {Cao}\ \emph {et~al.}(2000)\citenamefont {Cao},
  \citenamefont {Xu}, \citenamefont {Zhang}, \citenamefont {Chang},
  \citenamefont {Ho}, \citenamefont {Seelig}, \citenamefont {Liu},\ and\
  \citenamefont {Chang}}]{confinement2000randommedia}%
  \BibitemOpen
  \bibfield  {author} {\bibinfo {author} {\bibfnamefont {H.}~\bibnamefont
  {Cao}}, \bibinfo {author} {\bibfnamefont {J.~Y.}\ \bibnamefont {Xu}},
  \bibinfo {author} {\bibfnamefont {D.~Z.}\ \bibnamefont {Zhang}}, \bibinfo
  {author} {\bibfnamefont {S.-H.}\ \bibnamefont {Chang}}, \bibinfo {author}
  {\bibfnamefont {S.~T.}\ \bibnamefont {Ho}}, \bibinfo {author} {\bibfnamefont
  {E.~W.}\ \bibnamefont {Seelig}}, \bibinfo {author} {\bibfnamefont
  {X.}~\bibnamefont {Liu}},\ and\ \bibinfo {author} {\bibfnamefont {R.~P.~H.}\
  \bibnamefont {Chang}},\ }\bibfield  {title} {\bibinfo {title} {Spatial
  confinement of laser light in active random media},\ }\href
  {https://doi.org/10.1103/PhysRevLett.84.5584} {\bibfield  {journal} {\bibinfo
   {journal} {Phys. Rev. Lett.}\ }\textbf {\bibinfo {volume} {84}},\ \bibinfo
  {pages} {5584} (\bibinfo {year} {2000})}\BibitemShut {NoStop}%
\bibitem [{\citenamefont {Notomi}(2011)}]{confinement2011periodicity}%
  \BibitemOpen
  \bibfield  {author} {\bibinfo {author} {\bibfnamefont {M.}~\bibnamefont
  {Notomi}},\ }\bibfield  {title} {\bibinfo {title} {Strong light confinement
  with periodicity},\ }\href {https://doi.org/10.1109/JPROC.2011.2123850}
  {\bibfield  {journal} {\bibinfo  {journal} {Proc. IEEE}\ }\textbf {\bibinfo
  {volume} {99}},\ \bibinfo {pages} {1768} (\bibinfo {year}
  {2011})}\BibitemShut {NoStop}%
\bibitem [{\citenamefont {Riboli}\ \emph {et~al.}(2014)\citenamefont {Riboli},
  \citenamefont {Caselli}, \citenamefont {Vignolini}, \citenamefont {Intonti},
  \citenamefont {Vynck}, \citenamefont {Barthelemy}, \citenamefont {Gerardino},
  \citenamefont {Balet}, \citenamefont {Li}, \citenamefont {Fiore},
  \citenamefont {Gurioli},\ and\ \citenamefont
  {Wiersma}}]{confinement2014disorededmedia}%
  \BibitemOpen
  \bibfield  {author} {\bibinfo {author} {\bibfnamefont {F.}~\bibnamefont
  {Riboli}}, \bibinfo {author} {\bibfnamefont {N.}~\bibnamefont {Caselli}},
  \bibinfo {author} {\bibfnamefont {S.}~\bibnamefont {Vignolini}}, \bibinfo
  {author} {\bibfnamefont {F.}~\bibnamefont {Intonti}}, \bibinfo {author}
  {\bibfnamefont {K.}~\bibnamefont {Vynck}}, \bibinfo {author} {\bibfnamefont
  {P.}~\bibnamefont {Barthelemy}}, \bibinfo {author} {\bibfnamefont
  {A.}~\bibnamefont {Gerardino}}, \bibinfo {author} {\bibfnamefont
  {L.}~\bibnamefont {Balet}}, \bibinfo {author} {\bibfnamefont {L.~H.}\
  \bibnamefont {Li}}, \bibinfo {author} {\bibfnamefont {A.}~\bibnamefont
  {Fiore}}, \bibinfo {author} {\bibfnamefont {M.}~\bibnamefont {Gurioli}},\
  and\ \bibinfo {author} {\bibfnamefont {D.~S.}\ \bibnamefont {Wiersma}},\
  }\bibfield  {title} {\bibinfo {title} {Engineering of light confinement in
  strongly scattering disordered media},\ }\href
  {https://doi.org/10.1038/nmat3966} {\bibfield  {journal} {\bibinfo  {journal}
  {Nat. Mater}\ }\textbf {\bibinfo {volume} {13}},\ \bibinfo {pages} {720}
  (\bibinfo {year} {2014})}\BibitemShut {NoStop}%
\bibitem [{\citenamefont {Cooper}\ \emph {et~al.}(2013)\citenamefont {Cooper},
  \citenamefont {Wright}, \citenamefont {S\"{o}ller},\ and\ \citenamefont
  {Smith}}]{Cooper:13fockstates}%
  \BibitemOpen
  \bibfield  {author} {\bibinfo {author} {\bibfnamefont {M.}~\bibnamefont
  {Cooper}}, \bibinfo {author} {\bibfnamefont {L.~J.}\ \bibnamefont {Wright}},
  \bibinfo {author} {\bibfnamefont {C.}~\bibnamefont {S\"{o}ller}},\ and\
  \bibinfo {author} {\bibfnamefont {B.~J.}\ \bibnamefont {Smith}},\ }\bibfield
  {title} {\bibinfo {title} {Experimental generation of multi-photon {F}ock
  states},\ }\href {https://doi.org/10.1364/OE.21.005309} {\bibfield  {journal}
  {\bibinfo  {journal} {Opt. Express}\ }\textbf {\bibinfo {volume} {21}},\
  \bibinfo {pages} {5309} (\bibinfo {year} {2013})}\BibitemShut {NoStop}%
\bibitem [{\citenamefont {Chang}\ \emph {et~al.}(2014)\citenamefont {Chang},
  \citenamefont {Vuleti{\'c}},\ and\ \citenamefont
  {Lukin}}]{manybody2014luking}%
  \BibitemOpen
  \bibfield  {author} {\bibinfo {author} {\bibfnamefont {D.~E.}\ \bibnamefont
  {Chang}}, \bibinfo {author} {\bibfnamefont {V.}~\bibnamefont {Vuleti{\'c}}},\
  and\ \bibinfo {author} {\bibfnamefont {M.~D.}\ \bibnamefont {Lukin}},\
  }\bibfield  {title} {\bibinfo {title} {Quantum nonlinear optics---photon by
  photon},\ }\href {https://doi.org/10.1038/nphoton.2014.192} {\bibfield
  {journal} {\bibinfo  {journal} {Nat. Photon}\ }\textbf {\bibinfo {volume}
  {8}},\ \bibinfo {pages} {685} (\bibinfo {year} {2014})}\BibitemShut {NoStop}%
\bibitem [{\citenamefont {Pizzi}\ \emph {et~al.}(2023)\citenamefont {Pizzi},
  \citenamefont {Gorlach}, \citenamefont {Rivera}, \citenamefont {Nunnenkamp},\
  and\ \citenamefont {Kaminer}}]{emission2023manybody}%
  \BibitemOpen
  \bibfield  {author} {\bibinfo {author} {\bibfnamefont {A.}~\bibnamefont
  {Pizzi}}, \bibinfo {author} {\bibfnamefont {A.}~\bibnamefont {Gorlach}},
  \bibinfo {author} {\bibfnamefont {N.}~\bibnamefont {Rivera}}, \bibinfo
  {author} {\bibfnamefont {A.}~\bibnamefont {Nunnenkamp}},\ and\ \bibinfo
  {author} {\bibfnamefont {I.}~\bibnamefont {Kaminer}},\ }\bibfield  {title}
  {\bibinfo {title} {Light emission from strongly driven many-body systems},\
  }\href {https://doi.org/10.1038/s41567-022-01910-7} {\bibfield  {journal}
  {\bibinfo  {journal} {Nat. Phys}\ }\textbf {\bibinfo {volume} {19}},\
  \bibinfo {pages} {551} (\bibinfo {year} {2023})}\BibitemShut {NoStop}%
\bibitem [{\citenamefont {Guo}\ \emph {et~al.}(2020)\citenamefont {Guo},
  \citenamefont {Breum}, \citenamefont {Borregaard}, \citenamefont {Izumi},
  \citenamefont {Larsen}, \citenamefont {Gehring}, \citenamefont {Christandl},
  \citenamefont {Neergaard-Nielsen},\ and\ \citenamefont
  {Andersen}}]{distriubtesensing2020network}%
  \BibitemOpen
  \bibfield  {author} {\bibinfo {author} {\bibfnamefont {X.}~\bibnamefont
  {Guo}}, \bibinfo {author} {\bibfnamefont {C.~R.}\ \bibnamefont {Breum}},
  \bibinfo {author} {\bibfnamefont {J.}~\bibnamefont {Borregaard}}, \bibinfo
  {author} {\bibfnamefont {S.}~\bibnamefont {Izumi}}, \bibinfo {author}
  {\bibfnamefont {M.~V.}\ \bibnamefont {Larsen}}, \bibinfo {author}
  {\bibfnamefont {T.}~\bibnamefont {Gehring}}, \bibinfo {author} {\bibfnamefont
  {M.}~\bibnamefont {Christandl}}, \bibinfo {author} {\bibfnamefont {J.~S.}\
  \bibnamefont {Neergaard-Nielsen}},\ and\ \bibinfo {author} {\bibfnamefont
  {U.~L.}\ \bibnamefont {Andersen}},\ }\bibfield  {title} {\bibinfo {title}
  {Distributed quantum sensing in a continuous-variable entangled network},\
  }\href {https://doi.org/10.1038/s41567-019-0743-x} {\bibfield  {journal}
  {\bibinfo  {journal} {Nature Physics}\ }\textbf {\bibinfo {volume} {16}},\
  \bibinfo {pages} {281} (\bibinfo {year} {2020})}\BibitemShut {NoStop}%
\bibitem [{\citenamefont {Zhang}\ and\ \citenamefont
  {Zhuang}(2021)}]{Zhang_2021}%
  \BibitemOpen
  \bibfield  {author} {\bibinfo {author} {\bibfnamefont {Z.}~\bibnamefont
  {Zhang}}\ and\ \bibinfo {author} {\bibfnamefont {Q.}~\bibnamefont {Zhuang}},\
  }\bibfield  {title} {\bibinfo {title} {Distributed quantum sensing},\ }\href
  {https://doi.org/10.1088/2058-9565/abd4c3} {\bibfield  {journal} {\bibinfo
  {journal} {Quantum Sci. Technol.}\ }\textbf {\bibinfo {volume} {6}},\
  \bibinfo {pages} {043001} (\bibinfo {year} {2021})}\BibitemShut {NoStop}%
\bibitem [{\citenamefont {Pelayo}\ \emph {et~al.}(2023)\citenamefont {Pelayo},
  \citenamefont {Gietka},\ and\ \citenamefont {Busch}}]{gietka2023opticallatt}%
  \BibitemOpen
  \bibfield  {author} {\bibinfo {author} {\bibfnamefont {J.~C.}\ \bibnamefont
  {Pelayo}}, \bibinfo {author} {\bibfnamefont {K.}~\bibnamefont {Gietka}},\
  and\ \bibinfo {author} {\bibfnamefont {T.}~\bibnamefont {Busch}},\ }\bibfield
   {title} {\bibinfo {title} {Distributed quantum sensing with optical
  lattices},\ }\href {https://doi.org/10.1103/PhysRevA.107.033318} {\bibfield
  {journal} {\bibinfo  {journal} {Phys. Rev. A}\ }\textbf {\bibinfo {volume}
  {107}},\ \bibinfo {pages} {033318} (\bibinfo {year} {2023})}\BibitemShut
  {NoStop}%
\bibitem [{\citenamefont {Luca}\ and\ \citenamefont
  {Scardicchio}(2013)}]{DeLuca_2013ergodi}%
  \BibitemOpen
  \bibfield  {author} {\bibinfo {author} {\bibfnamefont {A.~D.}\ \bibnamefont
  {Luca}}\ and\ \bibinfo {author} {\bibfnamefont {A.}~\bibnamefont
  {Scardicchio}},\ }\bibfield  {title} {\bibinfo {title} {Ergodicity breaking
  in a model showing many-body localization},\ }\href
  {https://doi.org/10.1209/0295-5075/101/37003} {\bibfield  {journal} {\bibinfo
   {journal} {Europhys Lett.}\ }\textbf {\bibinfo {volume} {101}},\ \bibinfo
  {pages} {37003} (\bibinfo {year} {2013})}\BibitemShut {NoStop}%
\bibitem [{\citenamefont {Gietka}\ \emph {et~al.}(2019)\citenamefont {Gietka},
  \citenamefont {Chwede\ifmmode~\acute{n}\else \'{n}\fi{}czuk}, \citenamefont
  {Wasak},\ and\ \citenamefont {Piazza}}]{Gietka2019ergodic}%
  \BibitemOpen
  \bibfield  {author} {\bibinfo {author} {\bibfnamefont {K.}~\bibnamefont
  {Gietka}}, \bibinfo {author} {\bibfnamefont {J.}~\bibnamefont
  {Chwede\ifmmode~\acute{n}\else \'{n}\fi{}czuk}}, \bibinfo {author}
  {\bibfnamefont {T.}~\bibnamefont {Wasak}},\ and\ \bibinfo {author}
  {\bibfnamefont {F.}~\bibnamefont {Piazza}},\ }\bibfield  {title} {\bibinfo
  {title} {Multipartite entanglement dynamics in a regular-to-ergodic
  transition: Quantum fisher information approach},\ }\href
  {https://doi.org/10.1103/PhysRevB.99.064303} {\bibfield  {journal} {\bibinfo
  {journal} {Phys. Rev. B}\ }\textbf {\bibinfo {volume} {99}},\ \bibinfo
  {pages} {064303} (\bibinfo {year} {2019})}\BibitemShut {NoStop}%
\bibitem [{\citenamefont {Schulz}\ \emph {et~al.}(2019)\citenamefont {Schulz},
  \citenamefont {Hooley}, \citenamefont {Moessner},\ and\ \citenamefont
  {Pollmann}}]{STARKmbl2019pollmann}%
  \BibitemOpen
  \bibfield  {author} {\bibinfo {author} {\bibfnamefont {M.}~\bibnamefont
  {Schulz}}, \bibinfo {author} {\bibfnamefont {C.~A.}\ \bibnamefont {Hooley}},
  \bibinfo {author} {\bibfnamefont {R.}~\bibnamefont {Moessner}},\ and\
  \bibinfo {author} {\bibfnamefont {F.}~\bibnamefont {Pollmann}},\ }\bibfield
  {title} {\bibinfo {title} {Stark many-body localization},\ }\href
  {https://doi.org/10.1103/PhysRevLett.122.040606} {\bibfield  {journal}
  {\bibinfo  {journal} {Phys. Rev. Lett.}\ }\textbf {\bibinfo {volume} {122}},\
  \bibinfo {pages} {040606} (\bibinfo {year} {2019})}\BibitemShut {NoStop}%
\bibitem [{\citenamefont {Hafezi}\ \emph {et~al.}(2013)\citenamefont {Hafezi},
  \citenamefont {Mittal}, \citenamefont {Fan}, \citenamefont {Migdall},\ and\
  \citenamefont {Taylor}}]{Hafezi2013}%
  \BibitemOpen
  \bibfield  {author} {\bibinfo {author} {\bibfnamefont {M.}~\bibnamefont
  {Hafezi}}, \bibinfo {author} {\bibfnamefont {S.}~\bibnamefont {Mittal}},
  \bibinfo {author} {\bibfnamefont {J.}~\bibnamefont {Fan}}, \bibinfo {author}
  {\bibfnamefont {A.}~\bibnamefont {Migdall}},\ and\ \bibinfo {author}
  {\bibfnamefont {J.~M.}\ \bibnamefont {Taylor}},\ }\bibfield  {title}
  {\bibinfo {title} {Imaging topological edge states in silicon photonics},\
  }\href {https://doi.org/10.1038/nphoton.2013.274} {\bibfield  {journal}
  {\bibinfo  {journal} {Nat. Photon}\ }\textbf {\bibinfo {volume} {7}},\
  \bibinfo {pages} {1001–1005} (\bibinfo {year} {2013})}\BibitemShut
  {NoStop}%
\bibitem [{\citenamefont {Mittal}\ \emph {et~al.}(2014)\citenamefont {Mittal},
  \citenamefont {Fan}, \citenamefont {Faez}, \citenamefont {Migdall},
  \citenamefont {Taylor},\ and\ \citenamefont {Hafezi}}]{Mittal2014}%
  \BibitemOpen
  \bibfield  {author} {\bibinfo {author} {\bibfnamefont {S.}~\bibnamefont
  {Mittal}}, \bibinfo {author} {\bibfnamefont {J.}~\bibnamefont {Fan}},
  \bibinfo {author} {\bibfnamefont {S.}~\bibnamefont {Faez}}, \bibinfo {author}
  {\bibfnamefont {A.}~\bibnamefont {Migdall}}, \bibinfo {author} {\bibfnamefont
  {J.~M.}\ \bibnamefont {Taylor}},\ and\ \bibinfo {author} {\bibfnamefont
  {M.}~\bibnamefont {Hafezi}},\ }\bibfield  {title} {\bibinfo {title}
  {Topologically robust transport of photons in a synthetic gauge field},\
  }\href {https://doi.org/10.1103/PhysRevLett.113.087403} {\bibfield  {journal}
  {\bibinfo  {journal} {Phys. Rev. Lett.}\ }\textbf {\bibinfo {volume} {113}},\
  \bibinfo {pages} {087403} (\bibinfo {year} {2014})}\BibitemShut {NoStop}%
\bibitem [{\citenamefont {Gong}\ \emph {et~al.}(2018)\citenamefont {Gong},
  \citenamefont {Ashida}, \citenamefont {Kawabata}, \citenamefont {Takasan},
  \citenamefont {Higashikawa},\ and\ \citenamefont {Ueda}}]{Gong2018}%
  \BibitemOpen
  \bibfield  {author} {\bibinfo {author} {\bibfnamefont {Z.}~\bibnamefont
  {Gong}}, \bibinfo {author} {\bibfnamefont {Y.}~\bibnamefont {Ashida}},
  \bibinfo {author} {\bibfnamefont {K.}~\bibnamefont {Kawabata}}, \bibinfo
  {author} {\bibfnamefont {K.}~\bibnamefont {Takasan}}, \bibinfo {author}
  {\bibfnamefont {S.}~\bibnamefont {Higashikawa}},\ and\ \bibinfo {author}
  {\bibfnamefont {M.}~\bibnamefont {Ueda}},\ }\bibfield  {title} {\bibinfo
  {title} {Topological phases of non-hermitian systems},\ }\href
  {https://doi.org/10.1103/PhysRevX.8.031079} {\bibfield  {journal} {\bibinfo
  {journal} {Phys. Rev. X}\ }\textbf {\bibinfo {volume} {8}},\ \bibinfo {pages}
  {031079} (\bibinfo {year} {2018})}\BibitemShut {NoStop}%
\bibitem [{\citenamefont {Colella}\ \emph {et~al.}(2019)\citenamefont
  {Colella}, \citenamefont {Mivehvar}, \citenamefont {Piazza},\ and\
  \citenamefont {Ritsch}}]{Colella2019}%
  \BibitemOpen
  \bibfield  {author} {\bibinfo {author} {\bibfnamefont {E.}~\bibnamefont
  {Colella}}, \bibinfo {author} {\bibfnamefont {F.}~\bibnamefont {Mivehvar}},
  \bibinfo {author} {\bibfnamefont {F.}~\bibnamefont {Piazza}},\ and\ \bibinfo
  {author} {\bibfnamefont {H.}~\bibnamefont {Ritsch}},\ }\bibfield  {title}
  {\bibinfo {title} {Hofstadter butterfly in a cavity-induced dynamic synthetic
  magnetic field},\ }\href {https://doi.org/10.1103/PhysRevB.100.224306}
  {\bibfield  {journal} {\bibinfo  {journal} {Phys. Rev. B}\ }\textbf {\bibinfo
  {volume} {100}},\ \bibinfo {pages} {224306} (\bibinfo {year}
  {2019})}\BibitemShut {NoStop}%
\bibitem [{\citenamefont {Kosior}\ \emph {et~al.}()\citenamefont {Kosior},
  \citenamefont {Gietka}, \citenamefont {Mivehvar},\ and\ \citenamefont
  {Ritsch}}]{zenodo}%
  \BibitemOpen
  \bibfield  {author} {\bibinfo {author} {\bibfnamefont {A.}~\bibnamefont
  {Kosior}}, \bibinfo {author} {\bibfnamefont {K.}~\bibnamefont {Gietka}},
  \bibinfo {author} {\bibfnamefont {F.}~\bibnamefont {Mivehvar}},\ and\
  \bibinfo {author} {\bibfnamefont {H.}~\bibnamefont {Ritsch}},\ }\href
  {https://doi.org/https://doi.org/10.5281/zenodo.13304587} {\bibinfo {title}
  {Dynamical photon condensation into wannier-stark states}}\BibitemShut
  {NoStop}%
\bibitem [{\citenamefont {Claro}\ \emph {et~al.}(2003)\citenamefont {Claro},
  \citenamefont {Weisz},\ and\ \citenamefont {Curilef}}]{Claro2003}%
  \BibitemOpen
  \bibfield  {author} {\bibinfo {author} {\bibfnamefont {F.}~\bibnamefont
  {Claro}}, \bibinfo {author} {\bibfnamefont {J.~F.}\ \bibnamefont {Weisz}},\
  and\ \bibinfo {author} {\bibfnamefont {S.}~\bibnamefont {Curilef}},\
  }\bibfield  {title} {\bibinfo {title} {Interaction-induced oscillations in
  correlated electron transport},\ }\href
  {https://doi.org/10.1103/PhysRevB.67.193101} {\bibfield  {journal} {\bibinfo
  {journal} {Phys. Rev. B}\ }\textbf {\bibinfo {volume} {67}},\ \bibinfo
  {pages} {193101} (\bibinfo {year} {2003})}\BibitemShut {NoStop}%
\bibitem [{\citenamefont {Dias}\ \emph {et~al.}(2007)\citenamefont {Dias},
  \citenamefont {Nascimento}, \citenamefont {Lyra},\ and\ \citenamefont
  {de~Moura}}]{Dias2007}%
  \BibitemOpen
  \bibfield  {author} {\bibinfo {author} {\bibfnamefont {W.~S.}\ \bibnamefont
  {Dias}}, \bibinfo {author} {\bibfnamefont {E.~M.}\ \bibnamefont
  {Nascimento}}, \bibinfo {author} {\bibfnamefont {M.~L.}\ \bibnamefont
  {Lyra}},\ and\ \bibinfo {author} {\bibfnamefont {F.~A. B.~F.}\ \bibnamefont
  {de~Moura}},\ }\bibfield  {title} {\bibinfo {title} {Frequency doubling of
  {B}loch oscillations for interacting electrons in a static electric field},\
  }\href {https://doi.org/10.1103/PhysRevB.76.155124} {\bibfield  {journal}
  {\bibinfo  {journal} {Phys. Rev. B}\ }\textbf {\bibinfo {volume} {76}},\
  \bibinfo {pages} {155124} (\bibinfo {year} {2007})}\BibitemShut {NoStop}%
\bibitem [{\citenamefont {Khomeriki}\ \emph {et~al.}(2010)\citenamefont
  {Khomeriki}, \citenamefont {Krimer}, \citenamefont {Haque},\ and\
  \citenamefont {Flach}}]{Khomeriki2010}%
  \BibitemOpen
  \bibfield  {author} {\bibinfo {author} {\bibfnamefont {R.}~\bibnamefont
  {Khomeriki}}, \bibinfo {author} {\bibfnamefont {D.~O.}\ \bibnamefont
  {Krimer}}, \bibinfo {author} {\bibfnamefont {M.}~\bibnamefont {Haque}},\ and\
  \bibinfo {author} {\bibfnamefont {S.}~\bibnamefont {Flach}},\ }\bibfield
  {title} {\bibinfo {title} {Interaction-induced fractional {B}loch and
  tunneling oscillations},\ }\href {https://doi.org/10.1103/PhysRevA.81.065601}
  {\bibfield  {journal} {\bibinfo  {journal} {Phys. Rev. A}\ }\textbf {\bibinfo
  {volume} {81}},\ \bibinfo {pages} {065601} (\bibinfo {year}
  {2010})}\BibitemShut {NoStop}%
\bibitem [{\citenamefont {Wiater}\ \emph {et~al.}(2017)\citenamefont {Wiater},
  \citenamefont {Sowi\ifmmode~\acute{n}\else \'{n}\fi{}ski},\ and\
  \citenamefont {Zakrzewski}}]{Wiater2017}%
  \BibitemOpen
  \bibfield  {author} {\bibinfo {author} {\bibfnamefont {D.}~\bibnamefont
  {Wiater}}, \bibinfo {author} {\bibfnamefont {T.}~\bibnamefont
  {Sowi\ifmmode~\acute{n}\else \'{n}\fi{}ski}},\ and\ \bibinfo {author}
  {\bibfnamefont {J.}~\bibnamefont {Zakrzewski}},\ }\bibfield  {title}
  {\bibinfo {title} {Two bosonic quantum walkers in one-dimensional optical
  lattices},\ }\href {https://doi.org/10.1103/PhysRevA.96.043629} {\bibfield
  {journal} {\bibinfo  {journal} {Phys. Rev. A}\ }\textbf {\bibinfo {volume}
  {96}},\ \bibinfo {pages} {043629} (\bibinfo {year} {2017})}\BibitemShut
  {NoStop}%
\bibitem [{\citenamefont {Corrielli}\ \emph {et~al.}(2013)\citenamefont
  {Corrielli}, \citenamefont {Crespi}, \citenamefont {Della~Valle},
  \citenamefont {Longhi},\ and\ \citenamefont {Osellame}}]{Corrielli2013}%
  \BibitemOpen
  \bibfield  {author} {\bibinfo {author} {\bibfnamefont {G.}~\bibnamefont
  {Corrielli}}, \bibinfo {author} {\bibfnamefont {A.}~\bibnamefont {Crespi}},
  \bibinfo {author} {\bibfnamefont {G.}~\bibnamefont {Della~Valle}}, \bibinfo
  {author} {\bibfnamefont {S.}~\bibnamefont {Longhi}},\ and\ \bibinfo {author}
  {\bibfnamefont {R.}~\bibnamefont {Osellame}},\ }\bibfield  {title} {\bibinfo
  {title} {Fractional {B}loch oscillations in photonic lattices},\ }\href
  {https://doi.org/10.1038/ncomms2578} {\bibfield  {journal} {\bibinfo
  {journal} {Nat. Commun}\ }\textbf {\bibinfo {volume} {4}},\ \bibinfo {pages}
  {1555} (\bibinfo {year} {2013})}\BibitemShut {NoStop}%
\bibitem [{\citenamefont {Preiss}\ \emph {et~al.}(2015)\citenamefont {Preiss},
  \citenamefont {Ma}, \citenamefont {Tai}, \citenamefont {Lukin}, \citenamefont
  {Rispoli}, \citenamefont {Zupancic}, \citenamefont {Lahini}, \citenamefont
  {Islam},\ and\ \citenamefont {Greiner}}]{Preiss2015}%
  \BibitemOpen
  \bibfield  {author} {\bibinfo {author} {\bibfnamefont {P.~M.}\ \bibnamefont
  {Preiss}}, \bibinfo {author} {\bibfnamefont {R.}~\bibnamefont {Ma}}, \bibinfo
  {author} {\bibfnamefont {M.~E.}\ \bibnamefont {Tai}}, \bibinfo {author}
  {\bibfnamefont {A.}~\bibnamefont {Lukin}}, \bibinfo {author} {\bibfnamefont
  {M.}~\bibnamefont {Rispoli}}, \bibinfo {author} {\bibfnamefont
  {P.}~\bibnamefont {Zupancic}}, \bibinfo {author} {\bibfnamefont
  {Y.}~\bibnamefont {Lahini}}, \bibinfo {author} {\bibfnamefont
  {R.}~\bibnamefont {Islam}},\ and\ \bibinfo {author} {\bibfnamefont
  {M.}~\bibnamefont {Greiner}},\ }\bibfield  {title} {\bibinfo {title}
  {Strongly correlated quantum walks in optical lattices},\ }\href
  {https://doi.org/10.1126/science.1260364} {\bibfield  {journal} {\bibinfo
  {journal} {Science}\ }\textbf {\bibinfo {volume} {347}},\ \bibinfo {pages}
  {1229–1233} (\bibinfo {year} {2015})}\BibitemShut {NoStop}%
\bibitem [{\citenamefont {Kosior}\ \emph
  {et~al.}(2023{\natexlab{b}})\citenamefont {Kosior}, \citenamefont
  {Kokkelmans}, \citenamefont {Lewenstein}, \citenamefont {Zakrzewski},\ and\
  \citenamefont {P\l{}odzie\ifmmode~\acute{n}\else \'{n}\fi{}}}]{Kosior2023}%
  \BibitemOpen
  \bibfield  {author} {\bibinfo {author} {\bibfnamefont {A.}~\bibnamefont
  {Kosior}}, \bibinfo {author} {\bibfnamefont {S.}~\bibnamefont {Kokkelmans}},
  \bibinfo {author} {\bibfnamefont {M.}~\bibnamefont {Lewenstein}}, \bibinfo
  {author} {\bibfnamefont {J.}~\bibnamefont {Zakrzewski}},\ and\ \bibinfo
  {author} {\bibfnamefont {M.}~\bibnamefont {P\l{}odzie\ifmmode~\acute{n}\else
  \'{n}\fi{}}},\ }\bibfield  {title} {\bibinfo {title} {Phonon-assisted
  coherent transport of excitations in {R}ydberg-dressed atom arrays},\ }\href
  {https://doi.org/10.1103/PhysRevA.108.043308} {\bibfield  {journal} {\bibinfo
   {journal} {Phys. Rev. A}\ }\textbf {\bibinfo {volume} {108}},\ \bibinfo
  {pages} {043308} (\bibinfo {year} {2023}{\natexlab{b}})}\BibitemShut
  {NoStop}%
\bibitem [{\citenamefont {Bloch}\ \emph {et~al.}(2008)\citenamefont {Bloch},
  \citenamefont {Dalibard},\ and\ \citenamefont {Zwerger}}]{Bloch2008}%
  \BibitemOpen
  \bibfield  {author} {\bibinfo {author} {\bibfnamefont {I.}~\bibnamefont
  {Bloch}}, \bibinfo {author} {\bibfnamefont {J.}~\bibnamefont {Dalibard}},\
  and\ \bibinfo {author} {\bibfnamefont {W.}~\bibnamefont {Zwerger}},\
  }\bibfield  {title} {\bibinfo {title} {Many-body physics with ultracold
  gases},\ }\href {https://doi.org/10.1103/RevModPhys.80.885} {\bibfield
  {journal} {\bibinfo  {journal} {Rev. Mod. Phys.}\ }\textbf {\bibinfo {volume}
  {80}},\ \bibinfo {pages} {885} (\bibinfo {year} {2008})}\BibitemShut
  {NoStop}%
\end{thebibliography}
%


\widetext
\newpage
\setcounter{equation}{0}
\setcounter{figure}{0}
\renewcommand{\theequation}{S\arabic{equation}}
\renewcommand{\thefigure}{S\arabic{figure}}

\section{Supplemental Material}

\subsection{Derivation of the rotation frame Hamiltonian}

Consider an array of coupled Kerr resonators labeled by $j= -L/2, \ldots, L/2$. Each resonator in the array is assumed to have a distinct primary mode with a unique frequency denoted as $\omega_j$. The neighboring resonators are coupled through boson (e.g. photon) exchange, described as a nearest-neighbor tunneling process, where a boson leaks out of one resonator and enters the adjacent one. The full Hamiltonian in the lab frame is given by (see for example \cite{Jin2013, Sharma2021,Noh2016,Hartmann2016})
\begin{align}
\hat H_{\text{Lab}} = - J \sum_j \left(\hat a_j^\dagger \hat a_{j+1} + \text{H.c.}\right) + \sum_j \omega_j \hat a_j^\dagger \hat a_j + \chi \sum_j \hat a_j^{\dagger2} \hat a_j^2 + \sum_j \eta_j \left(\hat a_j e^{i \omega_p t} + \hat a_j^\dagger e^{-i \omega_p t}\right),
\end{align}
where $\hat a_j$, $\hat a_j^\dagger$ are bosonic operators for annihilating and creating bosons in the $j$-th resonator with frequency $\omega_j = \omega_0 + j\Delta \omega$. Here, $J$ is the tunneling amplitude, $\chi$ is the Kerr non-linearity leading to on-site interaction, and $\eta_j$ is the pumping rate of the $j$th resonator with a pumping laser of frequency $\omega_p$. 

To switch to a rotating frame with slowly varying variables, a unitary transformation with a time-dependent operator is applied
\begin{equation}
U(t) = \prod_j \exp\left(i \omega_p t \hat a_j^\dagger \hat a_j\right),
\end{equation}
transforming the bosonic operators to $\hat a'_j = U \hat a_j U^\dagger = \exp(-i\omega_p t) \hat a_j$ and the Hamiltonian to $\hat H = U \hat H_{\text{Lab}} U^\dagger + i \left(\partial_t U\right) U^\dagger$, i.e.,
\begin{align}\label{Ham_full}
\hat H = - J \sum_j \left(\hat a_j^\dagger \hat a_{j+1} + \text{H.c.}\right) + \sum_j \Delta_j \hat a_j^\dagger \hat a_j 
+ \chi \sum_j \hat a_j^{\dagger2} \hat a_j^2 + \sum_j \eta_j \left(\hat a_j + \hat a_j^\dagger\right),
\end{align}
where $\Delta_j = \Delta \omega \left(j - j_0\right)$ with $j_0 = (\omega_p - \omega_0) / \Delta \omega$.


\subsection{Bloch oscillations and consequences of U(1) symmetry}

Here, for pedagogical reasons, we first consider a non-interacting bosonic Hamiltonian in a tilted potential and show the system exhibits Bloch oscillations. Then, we introduce a pumping term $\sum_j \eta_j \left(\hat a_j + \hat a_j^\dagger\right)$ and discuss the consequences of U(1) symmetry breaking in weakly interacting systems. 

\subsubsection{Bloch oscillations}

The non-interacting Hamiltonian of our interest reads 
\be
\label{H_0_with_U1}
\hat H_0 = - J \sum_j \left(\hat a_j^\dagger \hat a_{j+1} + \text{H.c.} \right) +\sum_j \Delta_j \hat a_j^\dagger \hat a_j 
\ee
with $\Delta_j = \Delta \omega \left( j - j_0 \right) $, where the choice of $j_0$ is irrelevant due to the conservation of the particle number. 
The Hamiltonian is quadratic in $\hat a_j^\dagger$, $\hat a_{j}$ and can be readily diagonalised by a unitary transformation 
\be
\hat b_n = \sum_{j = -\infty}^\infty \beta_{n,j}\hat a_j ,
\quad \beta_{n,j} = 
\mathcal{J}_{j-n}\left(\gamma = \frac{2J}{\Delta \omega}\right) 
\ee
with $\mathcal{J}_{m}$ being the Bessel function of the first kind of order $m$. Since the transformation is unitary, then $\beta_{n,j}$ must fulfill 
\be\sum_j \beta_{n,j} \beta_{m,j} = \delta_{n,m}.
\ee
Indeed, using the Bessel function addition theorem 
\be
\sum_m \mathcal J_{m}(x) \mathcal J_{n-m}(y) = \mathcal J_{n}(x+y) 
\ee
as well as the property $\mathcal J_n(x) = \mathcal J_{-n}(-x) $, we obtain 
\be
\sum_j \mathcal J_{j-n}(\gamma) \mathcal J_{j-m}(\gamma) = \mathcal J_{m-n}(0) = \delta_{n,m}.
\ee
The inverse transformation then reads
\be
\hat a_j = \sum_n \beta_{n,j} \hat b_n .
\ee

\begin{figure}[t!]
 \centering
\includegraphics[width=1\textwidth]{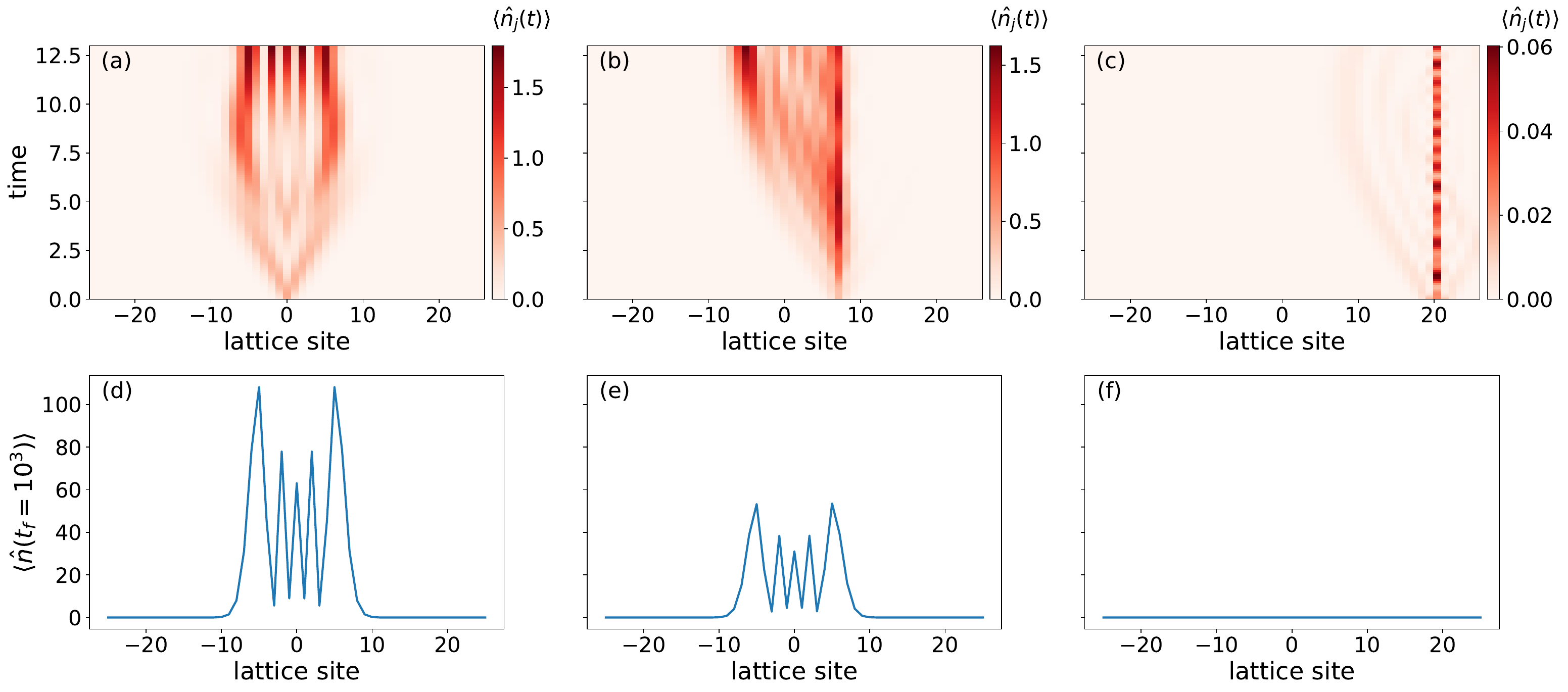}
\caption{
(a)-(c) The expectation values of particle number operators $\langle \hat n _j (t) \rangle$ in early times of evolution in a very weakly 
interacting case with a different pumping $\eta_j = \delta_{j,j'}$: 
(a) $j' = 0$,
(b) $j' = 7$,
(c) $j' = 20$. (d)-(e) The respective site occupation distributions after long time evolution in stationary states. In all panels $j_0=0$, $\Delta \omega = 0.3$ and $\chi = 10^{-2}$.  
}\label{fig_suppl3}
\end{figure}

The eigenstates $|\Psi_n \rangle = \hat b_n^\dagger |0\rangle $ are called the Wannier-Stark (WS) states. In practise, the infinite summation can be truncated, as the Bessel function $\mathcal{J}_{j-n}(\gamma) $ is mainly localised in the interval $ |j-n|< |\gamma|$.  
Although the formula holds for an infinite lattice, the WS states have finite spatial size, and therefore it can be used with a very good accuracy to describe bulk states of a finite-size lattice. Using the WS basis, the Hamiltonian \eqref{H_0_with_U1} can be recast into a diagonal form, 
\be
\hat H_0 =  \Delta \omega  \sum_n n\, \hat b_n^\dagger \hat b_n .
\ee 

The time evolution of the Wannier state, $|j\rangle = \hat a_j^\dagger |0\rangle$, reads
\be
|j (t) \rangle = \exp(i \hat H_0 \, t )|j\rangle = \sum_n \beta_{n,j} e^{i n \Delta \omega t } | \Psi_n \rangle
= \sum_{n,k} \beta_{n,j} \beta_{n,k} e^{i n \Delta \omega t } | k \rangle , 
\ee
which is a $T$-periodic function with period $T=2\pi/\Delta \omega$. Consequently, a state which is initially maximally localised on a tilted lattice performs Bloch oscillations around the initial position. 

\subsubsection{U(1) symmetry}

Because the total number of particles in the Hamiltonian $\hat H_0$ of Eq.~\eqref{H_0_with_U1} is conserved, we were able to neglect the term $ \hat K = - j_0 \sum_j \hat a_j^\dagger \hat a_j $, that commutes with the Hamiltonian. Equivalently, $\hat K$ can be gauged out by applying the time-dependent transformation 
\be
\label{U1_tr}
V(t) = \Pi_j \exp \left( i \chi(t) \hat a_j^\dagger \hat a_j \right), \quad \chi(t) = \int^t j_0(\tau) \mbox{d}\tau , 
\ee
where we assumed that $j_0$ can be explicitly time dependent. The transformation only modifies the global phase of the bosonic operators $\hat a'_j = V \hat a_j \tilde V^\dagger = \exp[-i \chi(t)] \hat a_j $ and does not contribute to the dynamics. It is a mere manifestation of the $U(1)$ symmetry of the model. 

The full Hamiltonian of our system \eqref{Ham_full}, which we study in the main text, consist of two additional terms, 
\be
\hat H = \hat H_0 + \chi \sum_j \hat a_j^{\dagger2} \hat a_j^2 + \sum_j \eta_j \left(\hat a_j + \hat a_j^\dagger\right),
\ee
where the first term describes contact boson-boson interactions, and the second accounts for coherent injection of bosons into the lattice. Although the interacting term is invariant under the $U(1)$ symmetry of Eq.~\eqref{U1_tr}, in general it is responsible for the decay of Bloch oscillations \cite{Buchleitner2003,Gustavsson2008,Krimer2009,Eckstein2011}. The second term explicitly breaks the $U(1)$ symmetry of the model, and therefore $ \hat K = - j_0 \sum_j \hat a_j^\dagger \hat a_j $ does not commute with $\hat H$ and cannot be neglected anymore. As described in the main text, the presence of $U(1)$ symmetry breaking terms in the Hamiltonian has enormous consequences to the system's dynamics and is responsible for the dynamical condensation. 

Additionally, in Fig.~\ref{fig_suppl3}(a)-(c) we show the early-time evolution of expectation values of particle number operators $\langle \hat n_j \rangle=\langle \hat a_j^\dag \hat a_j \rangle$ in each lattice site with different pumping schemes in the non-interacting case, while in Fig.~\ref{fig_suppl3}(d)-(e) we show the corresponding long-time occupation numbers of bosons over lattice sites. The final state is proportional to a central WS state irrespective of $j'$, as long as $j' $ lies within the WS localization range, i.e., approximately, $|j'| < |\gamma|$.

\subsection{Cumulant expansion}

In this section, we briefly introduce the cumulant expansion method. For an open system, the time evolution of an average of an observable $\hat{\mathcal{O}}$ can be calculated according to
\be
\frac{d \langle \hat{\mathcal{O}} \rangle }{dt} = \frac{i}{\hbar} \langle [ \hat H, \hat{\mathcal{O}} ] \rangle + \sum_j \kappa_j \langle 2 \hat c_j \hat{\mathcal{O}} \hat c_j^\dagger - \hat c_j^\dagger \hat c_j \hat{\mathcal{O}} - \hat{\mathcal{O}}  \hat c_j^\dagger \hat c_j \rangle, 
\ee
where $\kappa_j$ characterizes the $j$th decay channel and $c_j$ the corresponding jump operator. Writing down the equations explicitly, we typically find out an infinite hierarchy of equations for products of operators, which can be attributed to the non-commutativity of various operators. A well-established approach to deal with the infinite set of equations is to neglected quantum correlations. In such a case, an average value of operator product can be rewritten as product of average values of operators, $\langle \hat O_1 \hat O_2\rangle = \langle \hat O_1\rangle \langle \hat O_2\rangle$, which is typically referred to as the mean field approach. More generally, it is possible to neglect quantum correlations of an arbitrary order. A systematic approach to such a truncation is realized by a so-called cumulant expansion, which relies on decomposing average values of arbitrary operator products into products of average values of operators of a given lower order, leading to a closed set of equations.

\subsection{Interaction-induced Bloch oscillations}

\begin{figure}[bt]
 \centering
\includegraphics[width=0.45\textwidth]{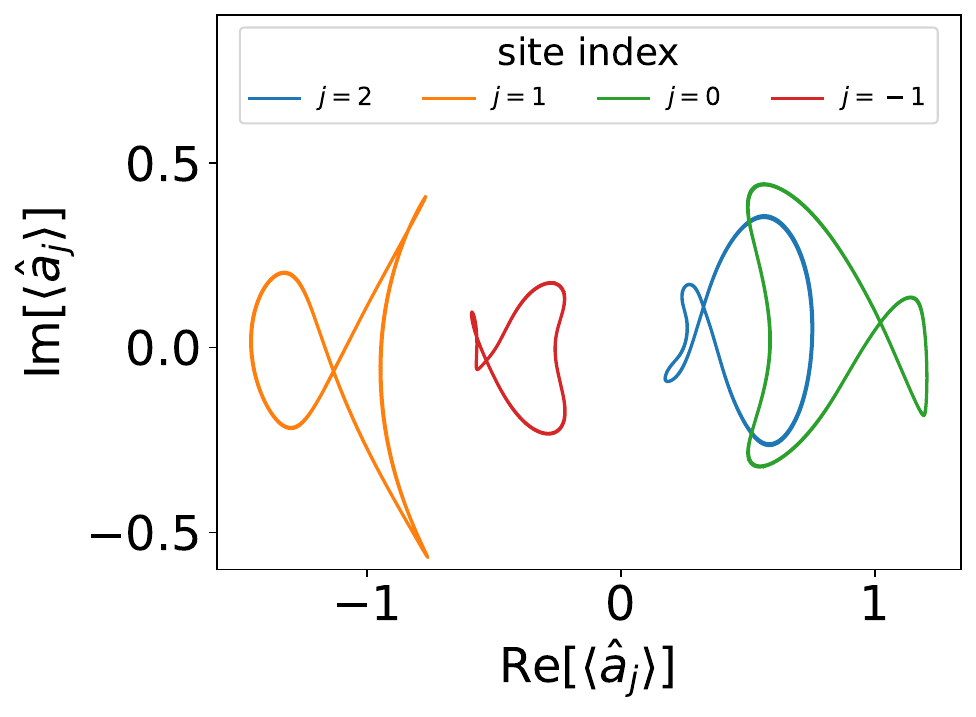}
\includegraphics[width=0.45\textwidth]{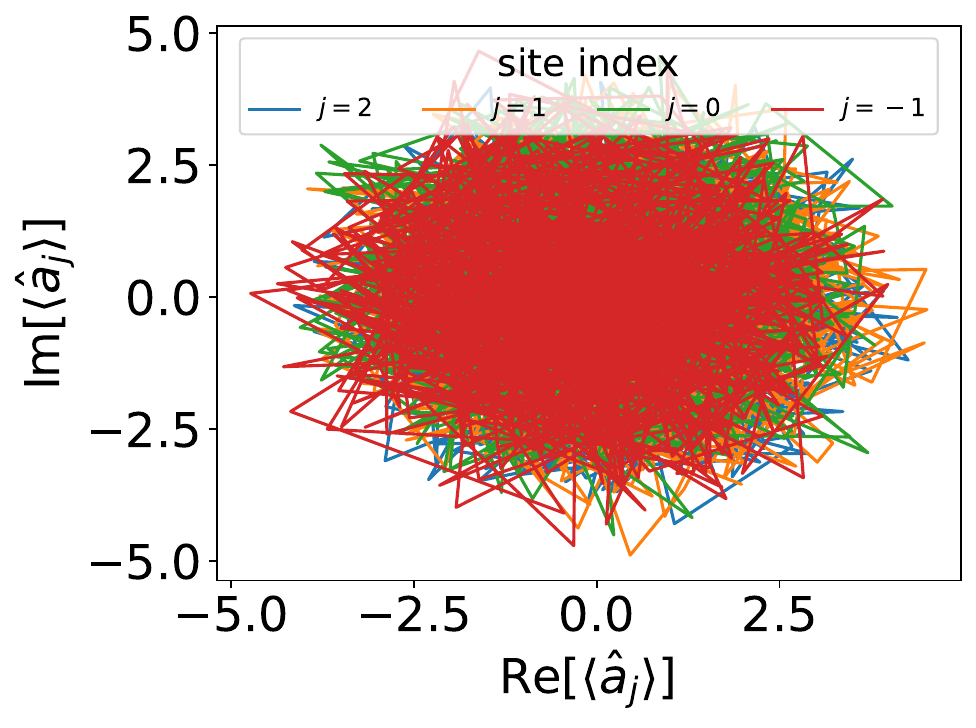}
\caption{
Evolution of the bosonic field amplitudes $\langle \hat a_j \rangle$ on the complex plane for $\Delta \omega = 0.5$ and (a) $\chi = 0.13$, (b) $\chi = 0.135$ [as in Fig.~4 in the main text]. (a) In the regime corresponding to the regular oscillatory behavior, the bosonic fields evolve along closed trajectories. (b) In the unstable regime, bosonic fields evolve chaotically. For the clarity of the presentation only selected modes are shown. 
}\label{fig_supp_LCs}
\end{figure}

\begin{figure}[bt]
 \centering
\includegraphics[width=1\textwidth]{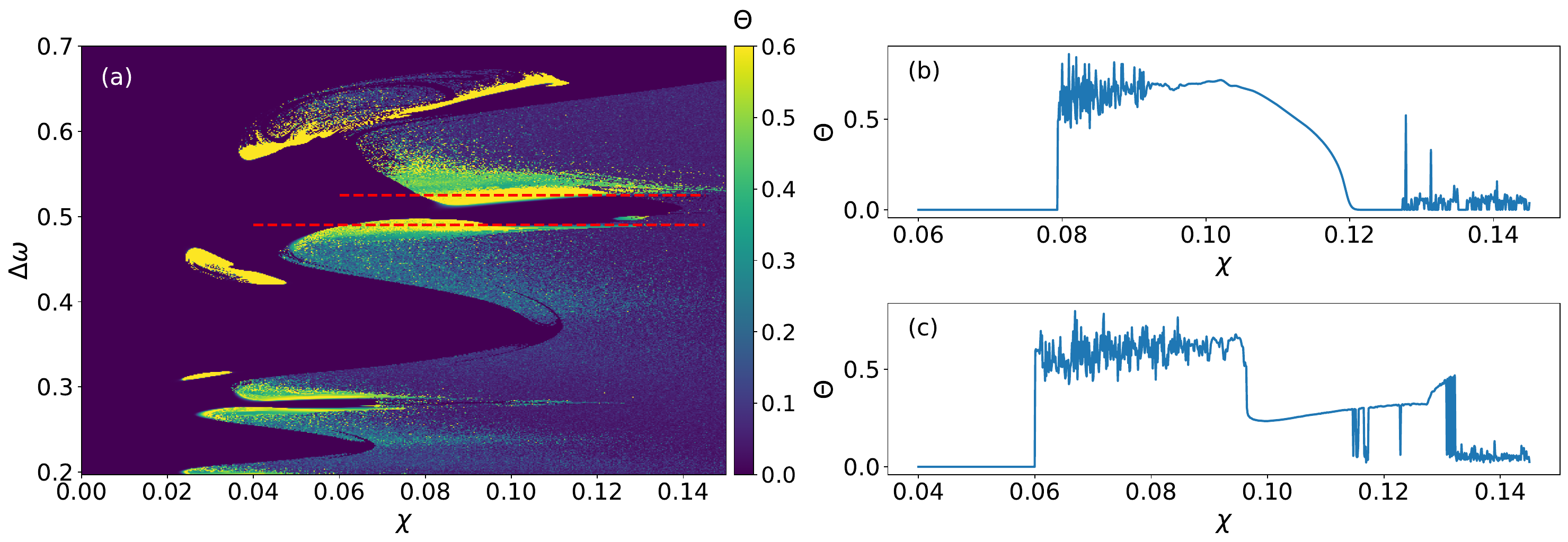}
\caption{The density plot of $\Theta$ which identifies stability islands of interaction-induced regular oscillations (bright colors). Panels (b) and (c) depict cuts through the density plot at the positions of red dashed lines. (b) $\Delta \omega = 0.525$, (c) $\Delta \omega = 0.49$. }\label{fig_suppl1}
\end{figure}

\begin{figure}[t]
 \centering
\includegraphics[width=1\textwidth]{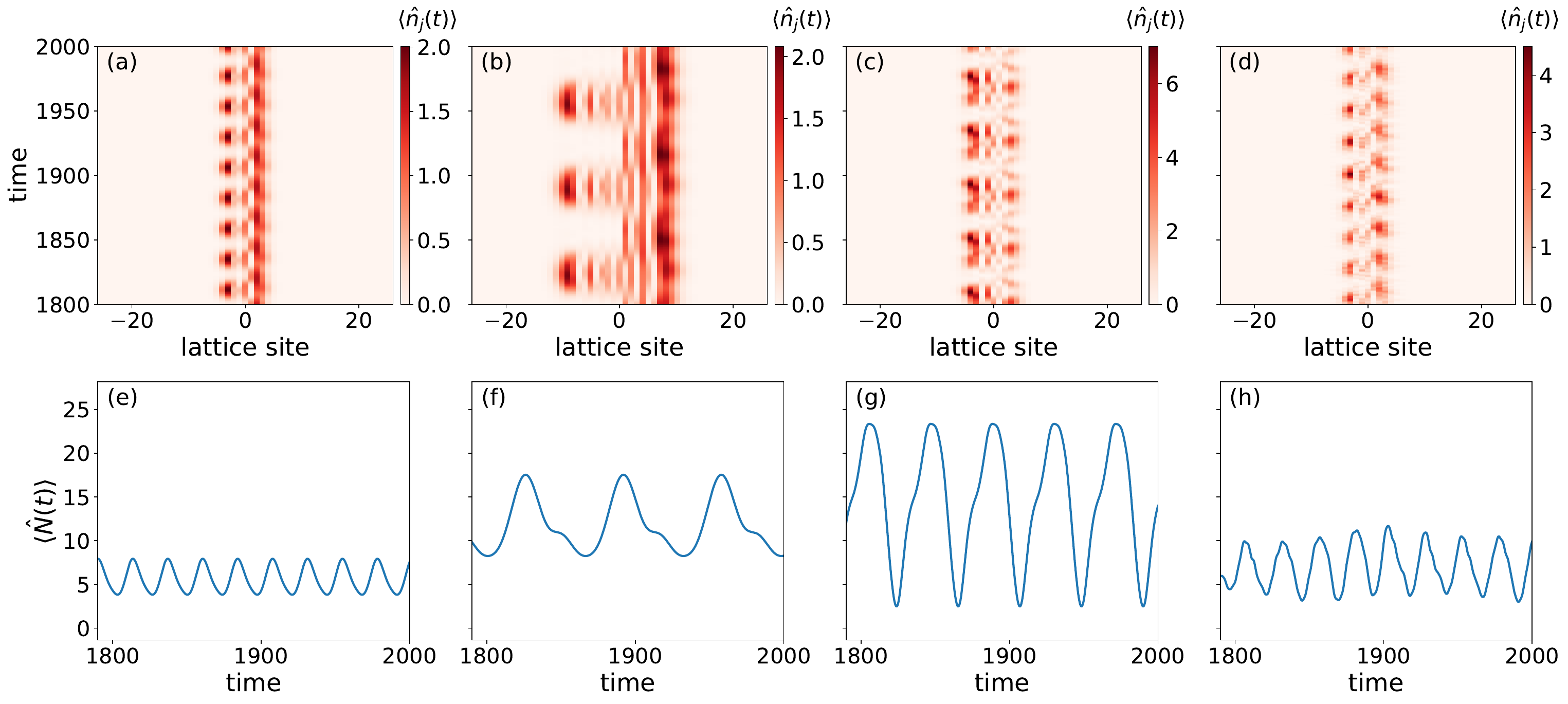}
\caption{
Panels present the expectation values of particle number operators $\langle \hat n _j (t) \rangle$ in the course of time evolution showing more examples of interaction-induced oscillations for different parameters of the system. 
(a) $\chi = 0.11$, $\Delta \omega = 0.525$
(b) $\chi = 0.05$, $\Delta \omega = 0.2$, 
(c) $\chi = 0.03$, $\Delta \omega = 0.45$, 
(d) $\chi = 0.09$, $\Delta \omega = 0.525$. Other parameters as in the main text of this Letter. 
}\label{fig_suppl2}
\end{figure}

In this section we focus on interaction-induced Bloch-like oscillations, which arise in our system whenever stationary solutions are not dynamically stable, in a similarity to the limit cycle solutions in driven dissipative systems [add citations]. Indeed, these oscillations can be interpreted as multi-mode limit cycles as shown in Fig.~\ref{fig_supp_LCs}(a). [On the other hand, stronger on-site interactions result in chaotic behavior, see Fig.~\ref{fig_supp_LCs}(b).] 

In the main text of this manuscript, we argue that the nonequilibrium phases of the system can be identified by two observables: 
\begin{itemize}
  \item $\Delta n$---relative change of the total boson number over a long-time evolution, defined as 
\be
\Delta n = \frac{\max_t \langle \hat N(t) \rangle - \min_t \langle \hat N(t) \rangle}{ \text{avg}_t \langle \hat N (t) \rangle},
\ee
with $\max_t $, $\min_t$ and $ \text{avg}_t $ denoting maximal, minimal and the average value of an observable during a long-time evolution (where the initial transient dynamics of the system has been neglected), and 
  \item $\langle \max[P_n (t)] \rangle_{av}$---time-averaged maximal fidelity (i.e. square modulus of the overlap) between the mean-field wavefunction $|\psi \rangle $ and the WS basis $|\Psi_n\rangle = \hat b_n^\dagger |0 \rangle = \sum_j \beta_{n,j} \hat a_j^\dagger |0\rangle$. 
\end{itemize}

In order to find the stable islands of interaction-induced Bloch oscillation it is convenient to define a third observable, as a simple product of the previous two observables,
\be
\Theta = \Delta n \cdot  \mathrm{avg}_t \langle \max[P_n (t)] \rangle
\ee
see Fig.~\ref{fig_suppl1}, where we show that $\Theta$ is close to zero both in the stationary and chaotic phases, and therefore it is a good observable to identify non-stationary regular dynamics in the system. 

\begin{figure}[bt]
 \centering
\includegraphics[width=0.5\textwidth]{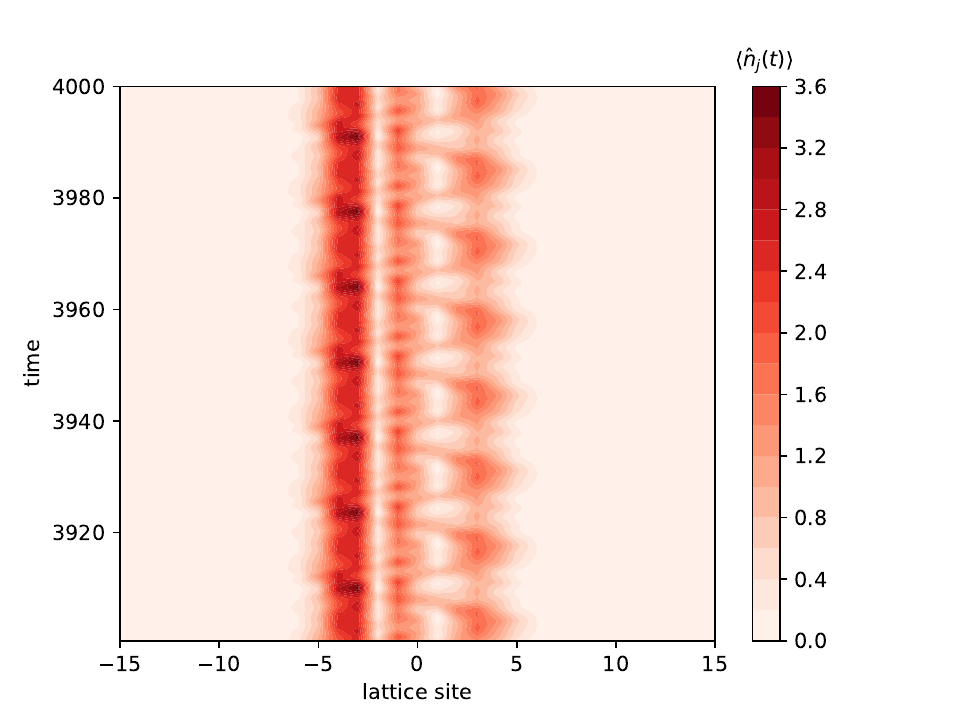}
\caption{
An example of oscillatory dynamics beyond the mean field in the second order cumulant expansion, for  $\chi = 0.03$, $\Delta \omega = 0.45$ (other parameters as in the main text of this Letter). 
}\label{fig_suppl_sec_order}
\end{figure}

In Fig.~\ref{fig_suppl2} we illustrate the expectation values of particle number operators $\langle \hat n _j (t) \rangle$ in the course of time evolution for different parameters of the system, showing more examples of interaction-induced oscillations.  Finally, in Fig.~\ref{fig_suppl_sec_order}, we present an example of oscillatory dynamics beyond the mean field in the second-order cumulant expansion. The panel clearly demonstrates that induced Bloch oscillations persist despite quantum fluctuations, though we anticipate quantitative differences in both the nature of the oscillations and the phase diagram

\subsection{Effect of next to nearest neighbor tunneling}
In this section we investigate the effect of including next-to-nearest neighbor (nnn) tunneling amplitudes $J_2$ in addition to the standard nearest neighbor tunneling $J$. Although in optical platforms the neighboring resonators are usually coupled evanescently, leading to negligible or nonexistent nnn tunneling amplitudes, the effect of nnn tunnelings could play a role in other platforms, such as optical lattice experiments. 

\begin{figure}[bt]
 \centering
\includegraphics[width=1\textwidth]{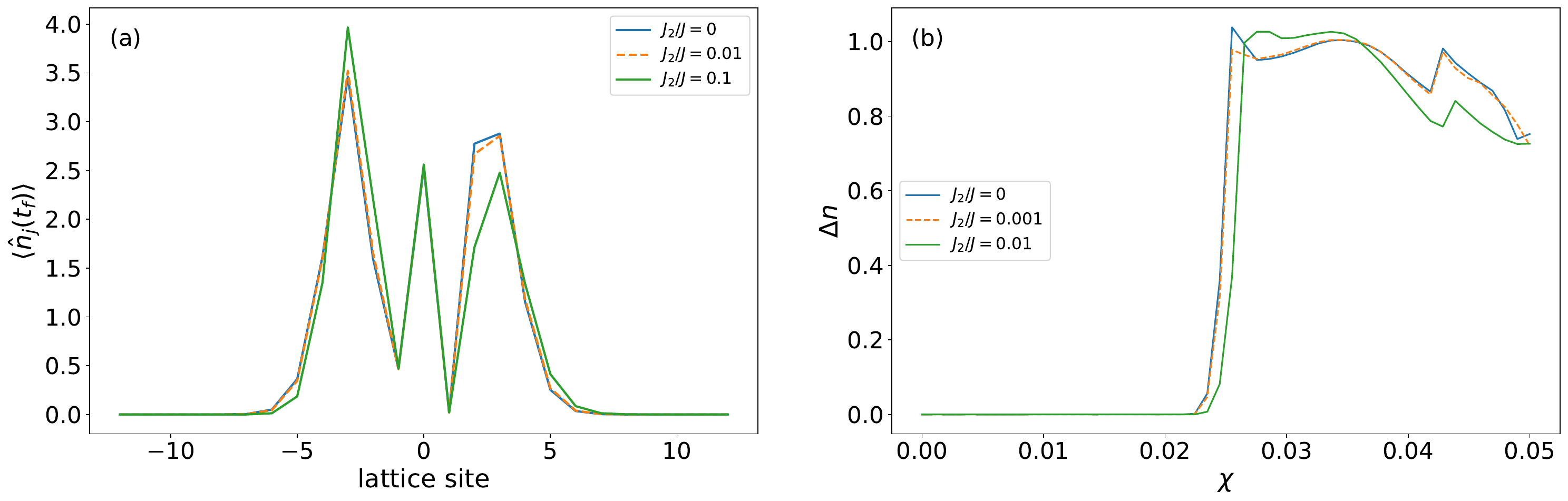}
\caption{
Effect of next to nearest neighbor (nnn) tunneling $J_2$. (a) Stationary state distribution with strong next-nearest-neighbor (nnn) tunneling amplitudes ($J_2 = 0.1J$), demonstrating that the qualitative nature of the distribution remains unchanged even for strong $J_2$. (b) Effects of stronger interactions on the stationary state distribution, showing that quantitative differences  can arise, particularly with a shift in the boundary between stationary and non-stationary regimes.
}\label{fig_suppl_nnn}
\end{figure}

In tight binding models, nnn tunnelings $J_2$ are generally much weaker than the dominant nearest neighbor tunnelings $J$ due to the extremely small overlap of localized Wannier states, causing the tunneling amplitudes to decay exponentially with the distance between lattice sites. To gain intuition, Table 1 of Ref. \cite{Bloch2008} shows that the $J_2/J$  ratio ranges between 0.1 and 0.01 for shallow optical lattice potentials and quickly drops for relatively deeper lattices. Figure~\ref{fig_suppl_nnn}(a) illustrates that even strong nnn tunneling amplitudes ($ J_2 = 0.1J $) do not qualitatively change the stationary state distribution. However, Figure~\ref{fig_suppl_nnn}(b) shows that there could be quantitative differences for stronger interactions, particularly in the shifting boundary between stationary and non-stationary regimes.

\end{document}